\documentclass[prb,aps,onecolumn,showpacs,floats,a4paper,times,12pt]{revtex4}
\usepackage{ulem}
\usepackage{cancel}
\usepackage{epsfig,bm,epsf,graphics}
\usepackage{xcolor}

\begin{document}

\title{Frustration - Exactly Solved Frustrated Models }
\author{H. T. Diep$^{a}$ and and H. Giacomini$^{b}$}
\address{$^{a}$ Laboratoire de Physique Th\'eorique et Mod\'elisation
Universit\'e de Cergy-Pontoise, CNRS, UMR 8089, 2 Avenue Adolphe
Chauvin, 95302 Cergy-Pontoise, Cedex, France.\\
 $^{b}$ Laboratoire de Math\'{e}matiques et Physique Th\'eorique
, Universit\'{e} de Tours, CNRS UMR 6083\\
Parc de Grandmont, 37200 Tours, France.}

\date{\today}

\begin{abstract}
After a short introduction on frustrated spin systems, we study in
this chapter several two-dimensional frustrated Ising spin systems
which can be exactly solved by using vertex models. We show that
these systems contain most of the spectacular effects due to the
frustration: high ground-state degeneracy, existence of several
phases in the ground-state phase diagram, multiple phase
transitions with increasing temperature, reentrance, disorder
lines, partial disorder at equilibrium. Evidences of such effects
in non solvable models are also shown and discussed.
\end{abstract}

\maketitle

\section{Frustration: an introduction}     

The study of order-disorder phenomena is a fundamental task of
equilibrium statistical mechanics. Great efforts have been made to
understand the basic mechanisms responsible for spontaneous
ordering as well as the nature of the phase transition in many
kinds of systems. In particular, during the last 25 years, much
attention has been paid to frustrated models.\cite {Lieb} The word
"frustration" has been introduced \cite {Tou,Villain1} to describe
the situation where a spin (or a number of spins) in the system
cannot find an orientation to {\it fully} satisfy all the
interactions with its neighboring spins (see below). This
definition can be applied to Ising spins, Potts models and vector
spins. In general, the frustration is caused either by competing
interactions (such as the Villain model\cite {Villain1}) or by
lattice structure as in the triangular, face-centered cubic (fcc)
and hexagonal-close-packed (hcp) lattices, with antiferromagnetic
nearest-neighbor (nn) interaction. The effects of frustration are
rich and often unexpected. Many of them are not understood yet at
present (see the other chapters of this book).

In addition to the fact that real magnetic materials are often
frustrated due to several kinds of interactions (see the chapter
by Gaulin and Gardner, this book), frustrated spin systems have
their own interest in statistical mechanics.  Recent studies show
that many established statistical methods and theories have
encountered many difficulties in dealing with frustrated systems.
In some sense, frustrated systems are excellent candidates to test
approximations and improve theories. Since the mechanisms of many
phenomena are not understood in real systems (disordered systems,
systems with long-range interaction, three-dimensional systems,
etc), it is worth to search for the origins of those phenomena in
exactly solved systems.  These exact results will help to
understand qualitatively the behavior of real systems which are in
general much more complicated.

\subsection{Definition}\index{frustration, definition}
Let us  give here some basic definitions to help readers
unfamiliar with these subjects to read the remaining chapters of
this book.

\begin{figure}[htb!] 
\centerline{\epsfig{file=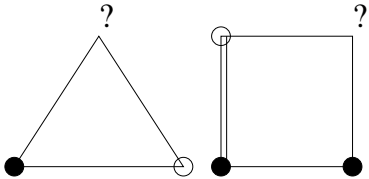,width=2in}}
\caption{Examples
of frustrated  plaquettes:  ferro- and antiferromagnetic
interactions, $J$ and $-J$, are shown by single and double lines,
$\uparrow$ and $\downarrow$ Ising spins by black and void circles,
respectively. Choosing any orientation for the spin marked by the
question mark will leave one of its bonds unsatisfied (frustrated
bond).} \label{fig:IntroFP}
\end{figure}

Consider two spins $\mathbf S_i$ and $\mathbf S_j$ with an
interaction $J$. The interaction energy is $E=-J \left(\mathbf S_i
\cdot \mathbf S_j\right)$. If $J$ is positive (ferromagnetic
interaction) then the minimum of $E$ is $-J$ corresponding to the
configuration in which $\mathbf S_i$ is parallel to $\mathbf S_j$.
If $J$ is negative (antiferromagnetic interaction), the minimum of
$E$ corresponds to the configuration where $\mathbf S_i$ is
antiparallel to $\mathbf S_j$.  It is easy to see that in a spin
system with nn ferromagnetic interaction, the ground state (GS) of
the system corresponds to the spin configuration where all spins
are parallel: the interaction of every pair  of spins is fully
satisfied. This is true for any lattice structure. If $J$ is
antiferromagnetic, the spin configuration of the GS depends on the
lattice structure: i) for lattices containing no elementary
triangles, i.e. bipartite lattices (such as square lattice, simple
cubic lattices, ...) the GS is the configuration in which each
spin is antiparallel to its neighbors, i.e. every bond is fully
satisfied. ii) for lattices containing elementary triangles such
as the triangular lattice, the fcc lattice and the hcp lattice,
one cannot construct a GS where all bonds are fully satisfied (see
Fig. \ref{fig:IntroFP}). The GS does not correspond to the minimum
of the interaction of every spin pair. In this case, one says that
the system is frustrated.

We consider another situation where the spin system can be
frustrated:
 this is the case with different kinds of conflicting interactions and the GS
does not correspond to the minimum of each kind of interaction.
For example, consider a chain of spins where the nn interaction
$J_1$ is ferromagnetic while the next nn (nnn) interaction $J_2$
is antiferromagnetic. As long as $|J_2|\ll J_1$, the GS is
ferromagnetic: every nn bond is then satisfied but the nnn ones
are not. Of course, when $|J_2|$ exceeds a critical value, the
ferromagnetic GS is no longer valid (see an example below): both
the nn and nnn bonds are not fully satisfied.

In a general manner, we can say that a spin system is frustrated
when one cannot find a configuration of spins to fully satisfy the
interaction (bond) between every pair of spins. In other words,
the minimum of the total energy does not correspond to the minimum
of each bond. This situation arises when there is a competition
between different kinds of interactions acting on a spin by its
neighbors or when the lattice geometry does not allow to satisfy
all the bonds simultaneously. With this definition, the chain with
nn ferromagnetic and nnn antiferromagnetic interactions discussed
above is frustrated even in the case where the ferromagnetic spin
configuration is its GS ($|J_2|\ll J_1$).

The first frustrated system which was studied in 1950 is the
triangular lattice with Ising spins interacting with each other
via a nn antiferromagnetic interaction\cite{Wan}. For vector
spins, non collinear spin configurations due to competing
interactions were first discovered in 1959 independently by
Yoshimori\cite{Yos}, Villain\cite{Vill} and Kaplan\cite{Kapl}.

Consider an elementary cell of the lattice. This cell is a polygon
formed by faces hereafter called "plaquettes". For example, the
elementary cell of the simple cubic lattice is a cube with six
square plaquettes, the elementary cell of the fcc lattice is a
tetrahedron formed by four triangular plaquettes. Let $J_{i,j}$ be
the interaction between two nn spins of the plaquette. According
to the definition of Toulouse,\cite{Tou} the plaquette is
frustrated if the parameter $P$ defined below is negative
\begin{equation}
P=\prod_{\left<i,j\right>}\mathrm{sign}(J_{i,j}), \label{frust1}
\end{equation}
where the product is performed over all $J_{i,j}$ around the
plaquette. Two examples of frustrated plaquettes are shown in Fig.
\ref{fig:IntroFP}: a triangle with three antiferromagnetic bonds
and a square with three ferromagnetic bonds and one
antiferromagnetic bond.  $P$ is negative in both cases.  One sees
that if one tries to put Ising spins on those plaquettes, at least
one of the bonds around the  plaquette will not be satisfied. For
vector spins, we show below that in the lowest energy state, each
bond is only partially satisfied.

One sees that for the triangular plaquette, the degeneracy is
three, and for the square plaquette it is four, in addition to the
degeneracy associated with returning all spins.  Therefore, the
degeneracy of an infinite lattice composed of such plaquettes is
infinite, in contrast to the unfrustrated case.

At this stage, we note that although  in the above discussion we
have taken the interaction between two spins to be of the form
$E=-J \left(\mathbf S_i \cdot \mathbf S_j\right)$, the concept of
frustration can be applied to other types of interactions such as
the Dzyaloshinski-Moriya interaction $E=-J \left|\left(\mathbf S_i
\wedge \mathbf S_j\right)\right|$: a spin system is frustrated
whenever the minimum of the system energy does not correspond to
the minimum of all local interactions, whatever the form of
interaction. We note however that this definition of frustration
is more general than the one using Eq. (\ref{frust1}).

The determination of the GS of various  frustrated Ising spin
systems as well as discussions on their properties will be shown .
In the following section, we analyze the GS of XY and Heisenberg
spins.

\subsection{Non collinear spin configurations}\index{non-collinear spin configuration}

Let us return to the plaquettes shown in Fig. \ref{fig:IntroFP}.
In the case of $XY$ spins, one can calculate the GS configuration
by minimizing the energy of the plaquette $E$ while keeping the
spin modulus constant. In the case of the triangular plaquette,
suppose that spin $\mathbf S_i$ $(i=1,2,3)$ of amplitude $S$ makes
an angle $\theta_i$ with the $\mathbf {Ox}$ axis. Writing $E$ and
minimizing it with respect to the angles $\theta_i$, one has
\begin{eqnarray}
E&=&J(\mathbf S_1\cdot \mathbf S_2+\mathbf S_2\cdot \mathbf S_3+\mathbf S_3\cdot \mathbf S_1)\nonumber \\
&=&JS^2\left[\cos (\theta_1-\theta_2)+\cos (\theta_2-\theta_3)+
\cos (\theta_3-\theta_1)\right],\nonumber \\
\frac{\partial E}{\partial \theta_1}&=&-JS^2\left[\sin
(\theta_1-\theta_2)-
\sin (\theta_3-\theta_1)\right]=0, \nonumber \\
\frac{\partial E}{\partial \theta_2}&=&-JS^2\left[\sin
(\theta_2-\theta_3)-
\sin (\theta_1-\theta_2)\right]=0, \nonumber \\
\frac{\partial E}{\partial \theta_3}&=&-JS^2\left[\sin
(\theta_3-\theta_1)- \sin (\theta_2-\theta_3)\right]=0. \nonumber
\end{eqnarray}

A solution of the last three equations is
$\theta_1-\theta_2=\theta_2 -\theta_3=\theta_3-\theta_1 =2\pi/3$.
One can also write
$$
E=J(\mathbf S_1\cdot \mathbf S_2+\mathbf S_2\cdot \mathbf
S_3+\mathbf S_3\cdot \mathbf S_1)
=-\frac{3}{2}JS^2+\frac{J}{2}(\mathbf S_1 + \mathbf S_2 + \mathbf
S_3)^2.
$$
The minimum here evidently corresponds to $\mathbf S_1 + \mathbf
S_2 + \mathbf S_3=0$ which yields the $120^\circ$
structure.\index{$120^\circ$ structure} This is true also for
Heisenberg spins.

We can do the same calculation for the case of the frustrated
square plaquette. Suppose that the antiferromagnetic bond connects
the spins $\mathbf{S}_1$ and $\mathbf{S}_2$. We find
\begin{equation}
\theta_2-\theta_1=\theta_3 -\theta_2=\theta_4-\theta_3
=\frac{\pi}{4} \textrm { and }\theta_1-\theta_4=\frac{3\pi}{4}
\label{frust2}
\end{equation}

If the antiferromagnetic bond is equal to $-\eta J$, the solution
for the angles is\cite{Berge}
\begin{equation}
\cos\theta_{32}=\cos\theta_{43}=\cos\theta_{14}\equiv \theta=\frac
{1}{2}[\frac {\eta+1}{\eta}]^{1/2}\label{frust2a}
\end{equation}
and $|\theta_{21}|=3|\theta|$, where $\cos\theta_{ij}\equiv
\cos\theta_{i}-\cos\theta_{j}$.

This solution exists if $| \cos\theta |\leq 1$, namely
$\eta>\eta_c=1/3$. One can check that when $\eta=1$, one has
$\theta =\pi/4$, $\theta_{21}=3\pi/4$.

\begin{figure}[htb!] 
\centerline{\epsfig{file=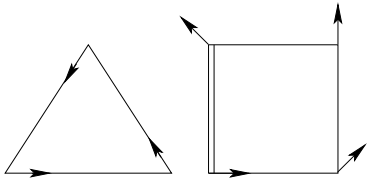,width=2.0in}} \caption{ Non
collinear spin configuration of frustrated triangular and square
plaquettes with $XY$ spins: ferro- and antiferromagnetic
interactions
 $J$ and $-J$ are indicated by thin and double lines, respectively.}
 \label{fig:IntroNCFT}
\end{figure}

We show the frustrated triangular and square lattices in Fig.
\ref{fig:IntroNCFT} with $XY$ spins ($N=2$).

One observes that there is  a two-fold degeneracy  resulting from
the symmetry by mirror reflecting with respect to an axis, for
example the $y$ axis in Fig. \ref{fig:IntroNCFT}. Therefore the
symmetry of these plaquettes is of Ising type O(1), in addition to
the symmetry SO(2) due to the invariance by global rotation of the
spins in the  plane. The lattices formed by these plaquettes will
be called in the following "antiferromagnetic triangular
lattice"\index{antiferromagnetic triangular lattice} and "Villain
lattice",\index{Villain lattice} respectively.

It is expected from the GS symmetry of these systems that the
transitions due to the respective breaking of O(1) and SO(2)
symmetries, if they occur at different temperatures, belongs
respectively to the 2D Ising universality class and to the
Kosterlitz-Thouless universality class. The question of whether
the two phase transitions would occur at the same temperature and
the nature of their universality remains at present an open
question. See more discussion in the chapter by Loison, this book.

The reader can find in Refs. [9] and [10] the derivation of the
non-trivial classical ground-state configuration of the fully
frustrated simple cubic lattice formed by stacking the
two-dimensional Villain lattices, in the case of Heisenberg and XY
spins.

\begin{figure}[htb!] 
\centerline{\epsfig{file=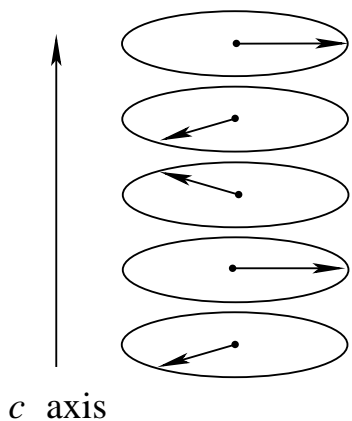,width=2in}} \caption{Helical
configuration when $\varepsilon = |J_2|/J_1>\varepsilon_c= 1/4$
($J_1>0$, $J_2<0$).}
 \label{fig:IntroHC}
\end{figure}

Another example is the case of a chain of Heisenberg spins with
ferromagnetic interaction $J_1(>0)$ between nn and
antiferromagnetic interaction
 $J_2 (<0)$ between nnn. When
$\varepsilon = |J_2|/J_1$ is larger than a critical value
$\varepsilon_c$, the spin configuration of the GS becomes non
collinear. One shows that the helical configuration displayed in
Fig. \ref{fig:IntroHC} is obtained by minimizing the interaction
energy:
\begin{eqnarray}
E&=&-J_1\sum_i\mathbf S_i\cdot\mathbf S_{i+1}
+|J_2|\sum_i\mathbf S_i\cdot\mathbf S_{i+2}\nonumber \\
&=&S^2\left[ -J_1\cos \theta+|J_2|\cos (2\theta)\right]\sum_i1\nonumber \\
\frac{\partial E}{\partial \theta}&=&S^2\left[J_1\sin \theta
-2|J_2|\sin(2\theta)\right]\sum_i1=0\nonumber \\
&=&S^2\left[J_1\sin \theta -4|J_2|\sin\theta \cos \theta
\right]\sum_i1=0,
\end{eqnarray}
where one has supposed that the angle between nn spins is
$\theta$.

The two solutions are
$$
\sin \theta=0 \longrightarrow \theta=0 \hspace{0.2cm}
\textrm{(ferromagnetic solution)}
$$
and
\begin{equation}
\cos \theta=\frac{J_1}{4|J_2|} \longrightarrow \theta= \pm \arccos
\left(\frac{J_1}{4|J_2|}\right).
\end{equation}
The last solution is possible if $-1\le \cos\theta \le 1$, i.e.
$J_1/\left(4|J_2|\right)\le 1$ or $|J_2|/J_1\ge 1/4 \equiv
\varepsilon_c$.

Again in this example, there are two degenerate configurations:
clockwise and counter-clockwise.

One defines in the following a chiral order parameter
\index{chiral order parameter} for each plaquette. For example, in
the case of a triangular plaquette, the chiral parameter is given
by
\begin{equation}\label{frust5}
\bm \kappa =\frac{2}{3\sqrt{3}}\left[\mathbf S_1\wedge \mathbf
S_2+\mathbf S_2\wedge \mathbf S_3+ \mathbf S_3\wedge \mathbf
S_1\right],
\end{equation}
where coefficient $2/\left(3\sqrt{3}\right)$  was introduced so
that the $\pm 2\pi/3$ degeneracy corresponds to $\kappa_i=\pm 1$.
\begin{figure}[htb!]  
\centerline{\epsfig{file=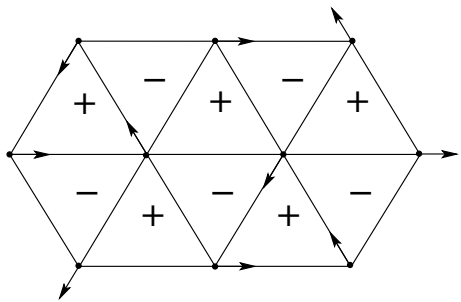,width=2.5in}}
\caption{\label{fig:IntroAFTL} Antiferromagnetic triangular
lattice with  $XY$ spins. The positive and negative chiralities
are indicated by $+$ and $-$.}
\end{figure}

We can form a triangular lattice using plaquettes as shown in Fig.
\ref{fig:IntroAFTL}. The GS corresponds to the state where all
plaquettes of the same orientation have the same chirality:
plaquettes $\bigtriangleup$ have positive chirality ($\kappa=1$)
and plaquettes $\bigtriangledown$ have negative chirality
($\kappa=-1$).  In terms of Ising spins, we have a perfect
antiferromagnetic order. This order is broken at a phase
transition temperature where $\kappa$ vanishes.

Let us enumerate two frequently encountered frustrated spin
systems where the nn interaction is antiferromagnetic: the fcc
lattice and the hcp lattice. These two lattices are formed by
stacking tetrahedra with four triangular faces.  The frustration
due to the lattice structure such as in these cases is called
"geometry frustration".

\section{Frustrated Ising spin systems}

We are interested here in frustrated Ising spin systems without
disorder. A review of early works (up to about 1985) on frustrated
Ising systems with periodic interactions, i.e. no bond disorder,
has been given by Liebmann.\cite {Lieb} These systems have their
own interest in statistical mechanics because they are
periodically defined and thus subject to exact treatment. To date,
very few systems are exactly solvable. They are limited to one and
two dimensions (2D).\cite {Bax} A few well-known systems showing
remarkable properties include the centered square lattice\cite
{Vaks} and its generalized versions,\cite{Mo,Chi}, the Kagom\'{e}
lattice,\cite {Ka/Na,Aza87,Diep91b} an anisotropic centered
honeycomb lattice,\cite{Diep91a} and several periodically dilute
centered square lattices.\cite{Diep92} Complicated cluster
models,\cite{Kita} and a particular three-dimensional case have
also been solved.\cite{Horiguchi} The phase diagrams in frustrated
models show a rich behavior. Let us mention a few remarkable
consequences of the frustration which are in connection with what
will be shown in this chapter. The degeneracy of the ground state
is very high, often infinite. At finite temperatures, in some
systems the degeneracy is reduced by thermal fluctuations which
select a number of states with largest entropy. This has been
called "Order by Disorder",\cite{Villain/al}\index{order by
disorder} in the Ising case. Quantum fluctuations and/or thermal
fluctuations can also select particular spin configurations in the
case of vector spins.\cite {Ogu,Hen} Another striking phenomenon
is the coexistence of Order and Disorder at equilibrium: a number
of spins in the system are disordered at all temperatures even in
an ordered phase.\cite {Aza87} The frustration is also at the
origin of the reentrance phenomenon. A reentrant phase can be
defined as a phase with no long-range order, or no order at all,
occurring in a region below an ordered phase on the temperature
scale.  In addition, the frustration can also give rise to
disorder lines\index{disorder lines} in the phase diagram of many
systems as will be shown below.

In this chapter, we confine ourselves to exactly solved Ising spin
systems that show remarkable features in the phase diagram such as
the reentrance, successive transitions, disorder lines and partial
disorder.  Other Ising systems are treated in the chapter by Nagai
et al.  Also, the reentrance in disordered systems such as spin
glasses is discussed in the chapter by Kawashima and Rieger.

The systems we consider in this chapter are periodically defined
(without bond disorder). The frustration due to competing
interactions will itself induce disorder in the spin orientations.
The results obtained can be applied to physical systems that can
be mapped into a spin language. The chapter is organized as
follows.  In the next section, we outline the method which allows
to calculate the partition function and the critical varieties of
2D Ising models without crossing interactions. In particular, we
show in detail the mapping of these models onto the 16- and
32-vertex models. We also explain a decimation method for finding
disorder solutions.\index{disorder solutions} The purpose of this
section is to give the reader enough mathematical details so that,
if he wishes, he can apply these techniques to 2D Ising models
with non-crossing interactions. In section 4, we shall apply the
results of section 3 in some systems which present remarkable
physical properties. The systems studied in section 4 contain most
of interesting features of the frustration: high ground state
degeneracy, reentrance, partial disorder, disorder lines,
successive phase transitions, and some  aspects of the
random-field Ising model. In section 4 we show some evidences of
reentrance and partial disorder found in three-dimensional systems
and in systems with spins other than the Ising model (Potts model,
classical vector spins, quantum spins). A discussion on the origin
of the reentrance phenomenon and concluding remarks are given in
section 5.

\section
{Mapping between Ising models and vertex models}\index{vertex
models}

The 2D Ising model with non-crossing
interactions is exactly soluble. The problem of finding the
partition function can be transformed in a free-fermion
model.If the lattice is a complicated one, the mathematical
problem to solve is very cumbersome.

For numerous two-dimensional Ising models with non-crossing
interactions, there exists another method, by far easier,
to find the exact partition function. This method consists
in mapping the model on a 16-vertex model or a 32-vertex
model. If the Ising model does not have crossing
interactions, the resulting vertex model will be exactly
soluble.
We will apply this method for finding the exact solution of
several Ising models in two-dimensional lattices with
non-crossing interactions.

Let us at first introduce the 16-vertex model and the 32-
vertex model, and the cases for which these models satisfy
the free-fermion condition.

\subsection{The 16-vertex model}

\begin{figure}[th]      
\centerline{\psfig{file=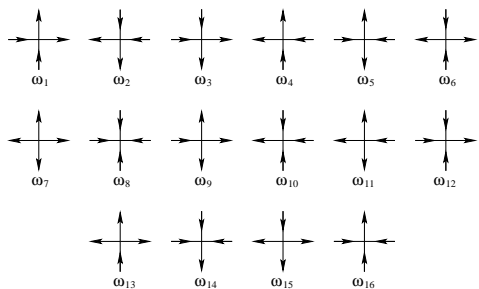,width=2.5in,angle=0}}
\vspace*{8pt} \caption{Arrow configurations and vertex weights of
the 16-vertex model. \label{re-fig5}}
\end{figure}

The 16-vertex model \index{16-vertex model} which we will consider
is a square lattice of N points, connected by edges between
neighboring sites. These edges can assume two states, symbolized
by right- and left- or up-and down-pointing arrows, respectively.
The allowed configurations of the system are characterized by
specifying the arrangement of arrows around each lattice point. In
characterizing these so-called vertex configurations, we follow
the enumeration of Baxter\cite{Bax}( see Fig.\ref{re-fig5}).

To each vertex we assign an energy $\epsilon_{k}( k = 1, 2, ..., 16)$
and a corresponding vertex weight (
Boltzmann factor) $ \omega_{k}=e^{\beta \epsilon_{k}}$
, where $\beta = (1)/(k_{B}T)$ , T being the temperature and
$k_{B}$ the
Boltzmann constant.
Then the partition function is
\begin{equation}
Z= \sum_{C} e^{-\beta (n_{1}\epsilon_{1}+...
+n_{16}\epsilon_{16})}\label{eq7}
\end{equation}
where the sum is over all allowed configurations C of arrows on
the lattice, $n_{j}$ is the number of vertex arrangements of type
j in configuration C. It is clear from Eq.(\ref{eq7}) that Z is a
function of the eight Boltzmann weights $ \omega_{k} ( k = 1, 2,
..., 16 )$ :
\begin{equation}
Z=Z(\omega_{1},...,\omega_{16})\label{eq8}
\end{equation}
So far, exact results have only been obtained for three
subclasses of the general 16-vertex model, i.e. the 6-vertex
( or ferroelectric ) model, the symmetric eight-vertex model
and the free-fermion model.\cite {Bax,Gaff}
Here we will consider only the case where the free-fermion
condition is satisfied, because in these cases the 16-vertex
model can be related to 2D Ising models without
crossing interactions.
Generally, a vertex model is soluble if the vertex weights
satisfy certain conditions so that the partition function is
reducible to the S matrix of a many-fermion system.\cite {Gaff} In
the present problem these constraints are the following :
\begin{eqnarray}
\omega_{1}&= &\omega_{2}\:,\: \omega_{3}=\omega_{4}  \nonumber \\
\omega_{5}&= &\omega_{6} \:,\: \omega_{7}=\omega_{8}  \nonumber \\
\omega_{9}&= &\omega_{10}=\omega_{11}=\omega_{12}  \nonumber \\
\omega_{13}&= &\omega_{14}=\omega_{15}=\omega_{16}  \nonumber \\
\omega_{1}\omega_{3}&+&\omega_{5}\omega_{7}
-\omega_{9}\omega_{11}-\omega_{13}\omega_{15}=0\label{re-eq9}
\end{eqnarray}

If these conditions are satisfied, the free energy of the
model can be expressed, in the thermodynamical limit, as
follows :
\begin{equation}
f=-\frac{1}{4\pi \beta}\ \int_{0}^{2\pi} d\phi \log \{A(\phi)
+[Q(\phi)]^{1/2}\}\label{eq10}
\end{equation}
where
\begin{eqnarray}
A(\phi)&= &a+c\cos (\phi)   \nonumber \\
Q(\phi)&= &y^{2}+z^{2}-x^{2}-2yz\cos (\phi)+x^{2}\cos ^{2}(\phi)  \nonumber \\
a&= &\frac{1}{2} ( \omega^{2}_{1}+\omega^{2}_{3}+2\omega_{1}\omega_{3}
+\omega^{2}_{5}+\omega^{2}_{7}+2\omega_{5}\omega_{7})
+2(\omega^{2}_{9}+\omega^{2}_{13})  \nonumber \\
c&= &2[\omega_{9}(\omega_{1}+\omega_{3})-\omega_{13}(\omega_{5}
+\omega_{7})]  \nonumber\\
y&= &2[\omega_{9}(\omega_{1}+\omega_{3})+\omega_{13}(\omega_{5}
+\omega_{7})]  \nonumber\\
z&= &\frac{1}{2} [(\omega_{1}+\omega_{3})^{2}-(\omega_{5}
+\omega_{7})^{2}]+2(\omega^{2}_{9}-\omega^{2}_{13})   \nonumber\\
x^{2}&= &z^{2}-\frac{1}{4}
[(\omega_{1}-\omega_{3})^{2}-(\omega_{5}
-\omega_{7})^{2}]^{2}\label{eq11}
\end{eqnarray}

Phase transitions occur when one or more pairs of zeros of the
expression $Q(\phi)$ close in on the real $\phi$ axis and "pinch"
the path of integration in the expression on the right-hand side
of Eq. (\ref{eq10}). This happens when $y^{2}=z^{2}$, i.e. when
\begin{equation}
\omega_{1}+\omega_{3}+\omega_{5}+\omega_{7}+
2\omega_{9}+2\omega_{13}=2\mbox{max}\{\omega_{1}+\omega_{3},
\omega_{5}+\omega_{7},2\omega_{9},2\omega_{13}\}\label{eq12}
\end{equation}

The type of singularity in the specific heat depends on
whether
\begin{eqnarray*}
(\omega_{1}-\omega_{3})^{2}-(\omega_{5}
-\omega_{7})^{2}\neq 0 \hspace{1.cm} (\mbox{logarithmic \, singularity})
\end{eqnarray*}
or
\begin{eqnarray}
(\omega_{1}-\omega_{3})^{2}-(\omega_{5} -\omega_{7})^{2} = 0
\hspace{1.cm} (\mbox{inverse\, square-root\,
singularity})\label{eq13}
\end{eqnarray}

\subsection{The 32-vertex model}

\begin{figure}[th]              
\centerline{\psfig{file=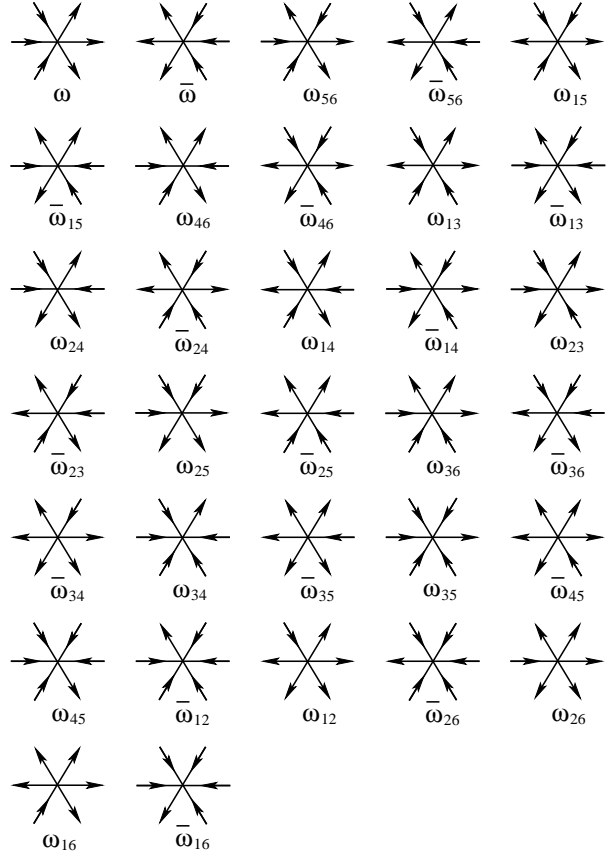,width=3.2in,angle=0}}
\vspace*{8pt} \caption{Arrow configurations and vertex weights of
the 32-vertex model. \label{re-fig6}}
\end{figure}

The 32-vertex model \index{32-vertex model} is defined by a
triangular lattice of N points, connected by edges between
neighboring sites. These edges can assume two states, symbolized
by an arrow pointing in or pointing out of a site. In the general
case, there are 64 allowed vertex configurations. If only an odd
number of arrows pointing into a site are allowed, we have 32
possible vertex configurations. This is the constraint that
characterizes the 32-vertex model. To each allowed vertex
configuration we assign an energy
 $\epsilon_{k}( k = 1, 2, ..., 32)$) and a corresponding
 vertex weight, defined
as it is shown in Fig. \ref{re-fig6}, where  $ \omega=e^{-\beta
\epsilon_{1}}$,
  $\overline{\omega}=e^{-\beta \epsilon_{2}}$,  $ \omega_{56}=
  e^{-\beta \epsilon_{3}}$,  $\overline{\omega}_{56}=
  e^{-\beta \epsilon_{4}}$
    ,  etc.

This notation for the Boltzmann vertex weights has been introduced
by Sacco and Wu,\cite{Sacco} and is used also by Baxter.\cite{Bax}
This model is not exactly soluble in the general case, but there
are several particular cases that are soluble.\cite {Sacco} Here
we will consider one of such cases, when the model satisfy the
free-fermion condition :\index{free-fermion condition}
\begin{eqnarray}
\omega\overline {\omega}&= &\omega_{12}\overline {\omega}_{12}-
\omega_{13}\overline {\omega}_{13}+\omega_{14}\overline {\omega}_{14}
-\omega_{15}\overline {\omega}_{15}+\omega_{16}\overline {\omega}_{16}
 \nonumber \\
\omega\omega_{mn}&= &\omega_{ij}\omega_{kl}
-\omega_{ik}\omega_{jl}+\omega_{il}\omega_{jk}\label{re-eq14}
\end{eqnarray}
for all permutations i, j, k, l, m, n   of   1, 2, ..., 6
such that $ m < n$  and $ i < j < k < l$.
There are 15 such permutations ( corresponding to the 15
choices of m and n ), and hence a total of 16
conditions.

The rather complicated notation for the Boltzmann weights is
justified by the condensed form of the free-fermion conditions Eq.
(\ref{re-eq14}).

When these conditions are satisfied, the free energy in the
thermodynamical limit can be expressed as
\begin{equation}
f=-\frac{1}{8\pi^{2} \beta}\ \int_{0}^{2\pi} d\theta
\int_{0}^{2\pi} d\phi \log [\omega^{2}
D(\theta,\phi)]\label{re-eq15}
\end{equation}
where

\begin{eqnarray}
\omega^{2}D(\theta,\phi)&= &\Omega_{1}^{2}+\Omega_{2}^{2}+
\Omega_{3}^{2}+\Omega_{4}^{2}-2(\Omega_{1}\Omega_{3}-
\Omega_{2}\Omega_{4})\cos (\theta) \nonumber \\
     & &-2(\Omega_{1}\Omega_{4}-\Omega_{2}\Omega_{3})\cos (\phi)
+2(\Omega_{3}\Omega_{4}-\Omega_{5}\Omega_{6})\cos (\theta+\phi)
 \nonumber \\
     & &+2(\Omega_{5}\Omega_{6}-\Omega_{1}\Omega_{2})\cos (\theta-\phi)
-4a\sin (\phi)\sin (\theta+\phi) \nonumber \\
     & &-4b\sin (\theta)\sin (\theta+\phi)
-2c\sin^{2}(\theta+\phi)-2d\sin^{2}(\theta)-2e\sin^{2}(\phi) \nonumber \\
& &\label{eq16}
\end{eqnarray}
with

\begin{eqnarray}
\Omega_{1}&= &\omega + \overline {\omega}\:,\:
\hspace{1.cm}\Omega_{2}=\omega_{25}
+ \overline {\omega}_{25} \nonumber \\
\Omega_{3}&= &\omega_{14} + \overline {\omega}_{14}\:,\:\hspace{1.cm}
\Omega_{4}=\omega_{36}
+ \overline {\omega}_{36} \nonumber \\
\Omega_{5}\Omega_{6}&= &\omega_{15}\omega_{24} +
\overline {\omega}_{15}
\overline {\omega}_{24}+\omega_{14}\overline {\omega}_{25}
+\omega_{25}\overline {\omega}_{14} \nonumber \\
a&= &\omega_{12}\omega_{45} + \overline {\omega}_{12}
\overline {\omega}_{45}-\omega\overline {\omega}_{36}
-\overline {\omega}\omega_{36} \nonumber \\
b&= &\omega_{23}\omega_{56} + \overline {\omega}_{23}
\overline {\omega}_{56}-\omega\overline {\omega}_{14}
-\overline {\omega}\omega_{14} \nonumber \\
c&= &\omega\overline{\omega} +\omega_{13}
\overline {\omega}_{13}-\omega_{12}\overline {\omega}_{12}
-\omega_{23}\overline{\omega}_{23} \nonumber\\
d&= &\omega\overline{\omega} +\omega_{26}
\overline {\omega}_{26}-\omega_{16}\overline {\omega}_{16}
-\omega_{12}\overline{\omega}_{12} \nonumber \\
e&= &\omega\overline{\omega} +\omega_{15} \overline
{\omega}_{15}-\omega_{56}\overline {\omega}_{56}
-\omega_{16}\overline{\omega}_{16}\label{eq17}
\end{eqnarray}

The critical temperature is determined  from the equation
\begin{equation}
\Omega_{1}+\Omega_{2}+\Omega_{3}+\Omega_{4}=
2\mbox{max}(\Omega_{1},\Omega_{2},\Omega_{3},\Omega_{4})\label{eq18}
\end{equation}

We will show now how different 2D Ising models
without crossing interactions can be mapped onto the
16-vertex model or the 32-vertex model, with the
free-fermion condition automatically satisfied in such
cases.

\begin{figure}[th]              
\centerline{\psfig{file=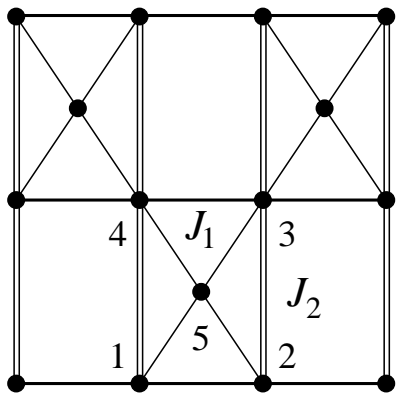,width=2.2in,angle=0}}
\vspace*{8pt} \caption{ Kagom\'{e} lattice. Interactions between
nearest neighbors and between next-nearest neighbors, $J_{1}$ and
$J_{2}$, are shown by single and double bonds, respectively.  The
lattice sites in a cell are numbered for decimation demonstration.
\label{re-fig7}}
\end{figure}

Let us consider at first an Ising model defined on a Kagom\'{e}
lattice, with two-spin interactions between nearest neighbors (nn)
and between next-nearest neighbors (nnn), $J_{1}$ and $J_{2}$,
respectively, as shown in Fig. \ref{re-fig7}.

The Hamiltonian is written as
\begin{equation}
H=-J_{1}\sum_{(ij)} \sigma_{i}\sigma_{j}-J_{2}\sum_{(ij)}
\sigma_{i}\sigma_{j}\label{re-eq19}
\end{equation}
where
and the
first and second sums run over the spin pairs
connected by single and double bonds, respectively.

The partition function is written as

\begin{equation}
Z=\sum_{\sigma} \prod_{c} \exp[K_{1}(\sigma_{1}\sigma_{5}+
\sigma_{2}\sigma_{5}+\sigma_{3}\sigma_{5}+
\sigma_{4}\sigma_{5}+\sigma_{1}\sigma_{2}+
\sigma_{3}\sigma_{4})+K_{2}(\sigma_{1}\sigma_{4}+
\sigma_{3}\sigma_{2})]\label{eq20}
\end{equation}
where $K_{1,2}=J_{1,2}/k_{B}T$  and where the sum is performed over all spin
configurations
and the product is taken over all elementary cells of the lattice.

\begin{figure}[th]              
\centerline{\psfig{file=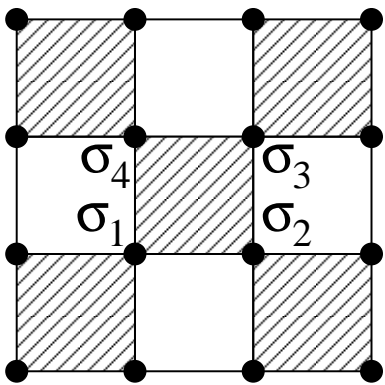,width=2.2in,angle=0}}
\vspace*{8pt} \caption{ The checkerboard lattice. At each shaded
square is associated the Boltzmann weight
$W(\sigma_{1},\sigma_{2},\sigma_{3},\sigma_{4})$, given in the
text. \label{re-fig8}}
\end{figure}

Since there are no crossing bond interactions, the system can be
transformed into an exactly solvable free-fermion model. We
decimate the central spin of each elementary cell of the lattice.
In doing so, we obtain a checkerboard Ising model with multispin
interactions (see Fig. \ref{re-fig8}).

The Boltzmann weight associated to each shaded square is
given by

\begin{eqnarray}
W(\sigma_{1},\sigma_{2},
\sigma_{3},\sigma_{4})&= &2\cosh (K_{1}(\sigma_{1}+\sigma_{2}+
\sigma_{3}+\sigma_{4}))\exp[K_{2}(\sigma_{1}\sigma_{4}+
\sigma_{2}\sigma_{3}) \nonumber \\
 & &+K_{1}(\sigma_{1}\sigma_{2}+
\sigma_{3}\sigma_{4})]\label{eq21}
\end{eqnarray}

The partition function of this checkerboard Ising model is
given by
\begin{equation}
Z=\sum_{\sigma} \prod W(\sigma_{1},\sigma_{2},
\sigma_{3},\sigma_{4})\label{eq22}
\end{equation}
where the sum is performed over all spin configurations and
the product is taken over all the shaded squares of the
lattice.

\begin{figure}[th]              
\centerline{\psfig{file=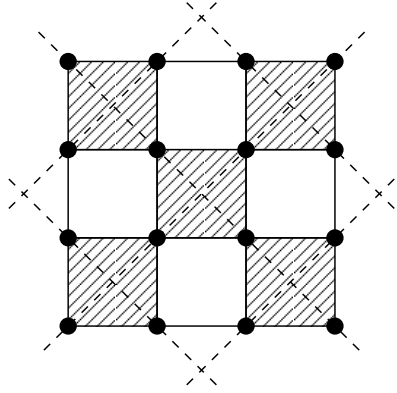,width=2.2in,angle=0}}
\vspace*{8pt} \caption{ The checkerboard lattice and the
associated square lattice with their bonds indicated by dashed
lines. \label{re-fig9}}
\end{figure}

In order to map this model onto the 16-vertex model, let us
introduce another square lattice where each site is placed at the
center of each shaded square of the checkerboard lattice, as shown
in Fig. \ref{re-fig9}.

\begin{figure}[th]              
\centerline{\psfig{file=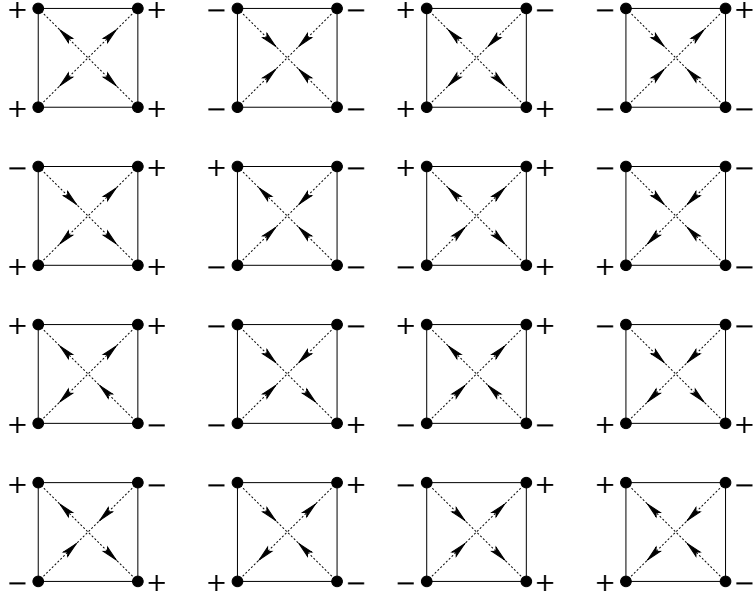,width=4.0in,angle=0}}
\vspace*{8pt} \caption{ The relation between spin configurations
and arrow configurations of the associated vertex model.
\label{re-fig10}}
\end{figure}

At each bond of this lattice we associate an arrow pointing out of
the site if the Ising spin that is traversed by this bond is equal
to +1, and pointing into the site if the Ising spin is equal to
-1, as it is shown in Fig. \ref{re-fig10}.

In this way, we have a 16-vertex model on the associated
square lattice. The Boltzmann weights of this vertex model
are expressed in terms of the Boltzmann weights of the
checkerboard Ising model, as follows
\begin{eqnarray}
\omega_{1}&= &W(-,-,+,+)\hspace{2.cm} \omega_{5}= W(-,+,-,+) \nonumber\\
\omega_{2}&= &W(+,+,-,-)\hspace{2.cm} \omega_{6}= W(+,-,+,-)\nonumber\\
\omega_{3}&= &W(-,+,+,-)\hspace{2.cm} \omega_{7}= W(+,+,+,+)\nonumber\\
\omega_{4}&= &W(+,-,-,+)\hspace{2.cm} \omega_{8}= W(-,-,-,-)\nonumber\\
\omega_{9}&= &W(-,+,+,+)\hspace{2.cm} \omega_{13}= W(+,-,+,+)\nonumber\\
\omega_{10}&= &W(+,-,-,-)\hspace{2.cm} \omega_{14}= W(-,+,-,-)\nonumber\\
\omega_{11}&= &W(+,+,-,+)\hspace{2.cm} \omega_{15}= W(+,+,+,-)\nonumber\\
\omega_{12}&= &W(-,-,+,-)\hspace{2.cm} \omega_{16}= W(-,-,-,+)\nonumber\\
& & \label{eq23}
\end{eqnarray}

Taking Eq. (\ref{eq21}) into account, we obtain
\begin{eqnarray}
\omega_{1}&=&\omega_{2} = 2e^{-2K_{2}+2K_{1}}\nonumber\\
\omega_{3}&=&\omega_{4} = 2e^{2K_{2}-2K_{1}}\nonumber\\
\omega_{5}&=&\omega_{6} = 2e^{-2K_{2}-2K_{1}}\nonumber\\
\omega_{7}&=&\omega_{8} = 2e^{2K_{2}+2K_{1}}\cosh (4K_{1})\nonumber\\
\omega_{9}&=&\omega_{10} =\omega_{11}=\omega_{12} =
\omega_{13}=\omega_{14} = \omega_{15}=\omega_{16} =2\cosh
(2K_{1})\label{re-eq24}
\end{eqnarray}

As can be easily verified, the free-fermion conditions Eq.
(\ref{re-eq9}) are identically satisfied by the Boltzmann weights
Eq. (\ref{re-eq24}), for arbitrary values of $K_{1}$ and $K_{2}$.
If we replace Eq. (\ref{re-eq24}) in Eq. (\ref{eq10}) and Eq.
(\ref{eq11}), we can obtain the explicit expression of the free
energy of the model. Moreover, by replacing Eq. (\ref{re-eq24}) in
Eq. (\ref{eq12}) we obtain the critical condition for this system
:

\begin{eqnarray}
\frac {1}{2}\ [\exp(2K_{1}+2K_{2})\cosh (4K_{1})+
\exp(-2K_{1}-2K_{2})]&+& \nonumber\\
\cosh (2K_{1}-2K_{2})+2\cosh (2K_{1})=
2\mbox{max}\{\frac {1}{2}\ [\exp(2K_{1}&+&2K_{2})\cosh (4K_{1})+ \nonumber\\
\exp(-2K_{1}-2K_{2})]\:;\:\cosh (2K_{2}-2K_{1})&;&\cosh
(2K_{1})\}\label{eq25}
\end{eqnarray}
which is decomposed into four critical lines depending on
the values of $J_{1}$ and $J_{2}$.

The singularity of the free energy is everywhere
logarithmic.

\begin{figure}[th]              
\centerline{\psfig{file=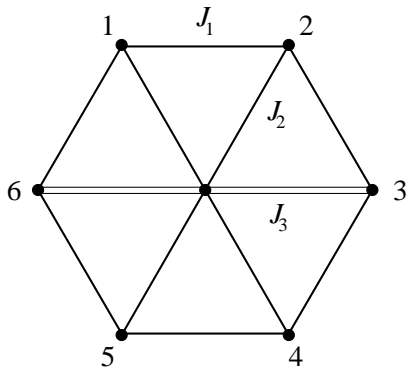,width=2.2in,angle=0}}
\vspace*{8pt} \caption{ Unit cell of the centered honeycomb
lattice: heavy, light, and double-light bonds denote the
interactions $J_{1}$, $J_{2}$, and $J_{3}$, respectively. The
sites on the honeycomb are numbered from 1 to 6 for decimation
demonstration (see text). \label{re-fig11}}
\end{figure}

Now, we will consider another 2D Ising model with two-spin
interactions and without crossing bonds. This model is defined on
a centered honeycomb lattice, as shown in Fig. \ref{re-fig11}.

The Hamiltonian of this model is as follows :

\begin{equation}
H=-J_{1}\sum_{(ij)} \sigma_{i}\sigma_{j}-J_{2}\sum_{(ij)}
\sigma_{i}\sigma_{j}-J_{3}
\sum_{(ij)}\sigma_{i}\sigma_{j}\label{eq26}
\end{equation}
where $\sigma_{i}=\pm 1$  is an Ising spin occupying the lattice
site i , and the first, second, and third sums run over the spin
pairs connected by heavy, light, and doubly light bonds,
respectively ( see Fig. \ref{re-fig11}). When $J_{2}=J_{3}=0$, one
recovers the honeycomb lattice, and when $J_{1}=J_{2}=J_{3}$,
one has the triangular lattice.

Let us denote the central spin in a lattice cell, shown in Fig.
\ref{re-fig11}, by $\sigma$, and number the other spins from
$\sigma_{1}$ to $\sigma_{6}$. The Boltzmann weight associated to
the elementary cell is given by

\begin{eqnarray*}
W = \exp[K_{1}(\sigma_{1}\sigma_{2}+
\sigma_{2}\sigma_{3}+\sigma_{3}\sigma_{4}+
\sigma_{4}\sigma_{5}+\sigma_{5}\sigma_{6}+
\sigma_{6}\sigma_{1})+
\end{eqnarray*}
\begin{equation}
K_{2}\sigma(\sigma_{1}+\sigma_{2}+\sigma_{4}+\sigma_{5})+
K_{3}\sigma(\sigma_{3}+\sigma_{6})]\label{eq27}
\end{equation}

The partition function of the model is written as

\begin{equation}
Z=\sum_{\sigma}\prod_{c} W\label{eq28}
\end{equation}
where the sum is performed over all spin configurations and the
product is taken over all elementary cells of the lattice.
Periodic boundary conditions are imposed. Since there is no
crossing-bond interaction, the model is exactly soluble. To obtain
the exact solution, we decimate the central spin of each
elementary cell of the lattice. In doing so, we obtain a honeycomb
Ising model with multispin interactions.

After decimation of each central spin, the Boltzmann factor
associated to an elementary cell is given by

\begin{eqnarray*}
W' = 2\exp[K_{1}(\sigma_{1}\sigma_{2}+
\sigma_{2}\sigma_{3}+\sigma_{3}\sigma_{4}+
\sigma_{4}\sigma_{5}+\sigma_{5}\sigma_{6}+
\sigma_{6}\sigma_{1})]\times
\end{eqnarray*}
\begin{equation}
\cosh [K_{2}(\sigma_{1}+\sigma_{2}+\sigma_{4}+\sigma_{5})+
K_{3}(\sigma_{3}+\sigma_{6})]\label{eq29}
\end{equation}

We will show in the following that this model is equivalent to a
special case of the 32-vertex model on the triangular lattice that
satisfies the free-fermion condition.

\begin{figure}[th]              
\centerline{\psfig{file=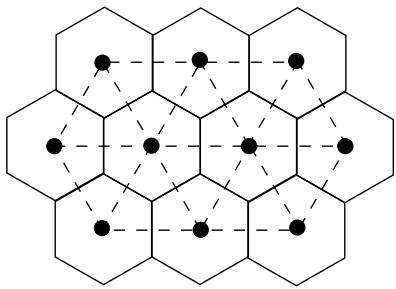,width=2.2in,angle=0}}
\vspace*{8pt} \caption{ The honeycomb lattice and the dual
triangular lattice, with their bonds indicated by dashed lines.
\label{re-fig12}}
\end{figure}

Let us consider the dual lattice of the honeycomb lattice, i.e.
the triangular lattice.\cite{Bax} The sites of the dual lattice
are placed at the center of each elementary cell and their bonds
are perpendicular to bonds of the honeycomb lattice, as it is
shown in Fig. \ref{re-fig12}.

\begin{figure}[th]              
\centerline{\psfig{file=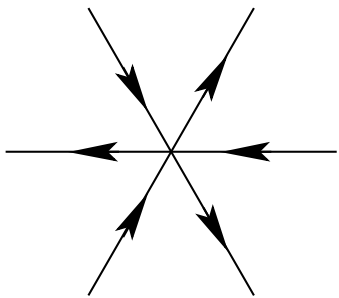,width=2in,angle=0}} \vspace*{8pt}
\caption{ The standard arrow configuration for the triangular
lattice. \label{re-fig13}}
\end{figure}

Each site of the triangular lattice is surrounded by 6 sites of
the honeycomb lattice. At each bond of the triangular lattice we
associate an arrow. We take the arrow configuration shown in Fig.
\ref{re-fig13} as the standard one. We can establish a two-to-one
correspondence between spin configurations of the honeycomb
lattice and arrow configurations in the triangular lattice. This
can be done in the following way : if the spins on either side of
a bond of the triangular lattice are equal ( different ), place an
arrow on the bond pointing in the same ( opposite ) way as the
standard. If we do this for all bonds, then at each site of the
triangular lattice there must be an even number of non-standard
arrows on the six incident bonds, and hence an odd number of
incoming ( and outgoing ) arrows. This is the property that
characterize the 32 vertex model on the triangular lattice.

\begin{figure}[th]              
\centerline{\psfig{file=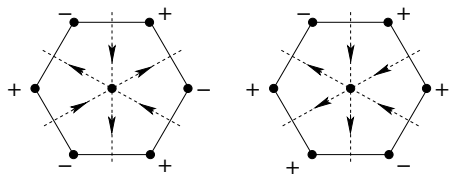,width=2.4in,angle=0}}
\vspace*{8pt} \caption{ Two cases of the correspondence between
arrow configurations and spin configurations. \label{re-fig14}}
\end{figure}

In Fig. \ref{re-fig14} we show two cases of the relation between
arrow configurations on the triangular lattice and spin
configurations on the honeycomb lattice.

In consequence, the Boltzmann weights of the 32-vertex model will
be a function of the Boltzmann weights $W'(\sigma_{1},\sigma_{2},
\sigma_{3},\sigma_{4},\sigma_{5},\sigma_{6})$ , associated to a
face of the honeycomb lattice. By using the relation between
vertex and spin configurations described above and expression Eq.
(\ref{eq29}), we find

\begin{eqnarray}
\omega&= &W'(+,-,-,-,+,+) = 2e^{2K_{1}} \nonumber\\
\overline {\omega}&= & W'(+,+,-,+,+,-) = 2e^{-2K_{1}}\cosh (4K_{2}
-2K_{3})\nonumber\\
\omega_{56}&= & W'(+,-,+,-,+,+) = 2e^{-2K_{1}}\cosh (2K_{3})\nonumber\\
\overline {\omega}_{56}&= & W'(+,+,+,+,+,-) =
2e^{2K_{1}}\cosh (4K_{2})\nonumber\\
\omega_{15}&= & W'(+,+,+,-,+,+) = 2e^{2K_{1}}\cosh (2K_{2}+
2K_{3})\nonumber\\
\overline {\omega}_{15}&= & W'(+,-,+,+,+,-) =
2e^{-2K_{1}}\cosh (2K_{2})\nonumber\\
\omega_{46}&= & W'(+,-,+,+,+,+) = 2e^{2K_{1}}\cosh (2K_{2}+
2K_{3})\nonumber\\
\overline {\omega}_{46}&= & W'(+,+,+,-,+,-) =
2e^{-2K_{1}}\cosh (2K_{2})\nonumber\\
\omega_{13}&= & W'(+,+,+,+,-,+) = 2e^{2K_{1}}\cosh (2K_{2}+
2K_{3})\nonumber\\
\overline {\omega}_{13}&= & W'(+,-,+,-,-,-) =
2e^{-2K_{1}}\cosh (2K_{2})\nonumber\\
\omega_{24}&= & W'(+,-,-,-,-,-) = 2e^{2K_{1}}\cosh (2K_{2}+
2K_{3})\nonumber\\
\overline {\omega}_{24}&= & W'(+,+,-,+,-,+) =
2e^{-2K_{1}}\cosh (2K_{2})\nonumber\\
\omega_{14}&= & W'(+,+,+,+,+,+) = 2e^{6K_{1}}\cosh (4K_{2}+
2K_{3})\nonumber\\
\overline {\omega}_{14}&= & W'(+,-,+,-,+,-) = 2e^{-6K_{1}}\nonumber\\
\omega_{23}&= & W'(+,-,-,-,+,-) = 2e^{-2K_{1}}\cosh (2K_{3})\nonumber\\
\overline {\omega}_{23}&= & W'(+,+,-,+,+,+) = 2e^{2K_{1}}\cosh (4K_{2})\nonumber\\
\omega_{25}&= & W'(+,-,-,+,-,-) = 2e^{-2K_{1}}\cosh (2K_{3})\nonumber\\
\overline {\omega}_{25}&= & W'(+,+,-,-,-,+) = 2e^{2K_{1}}\nonumber\\
\omega_{36}&= & W'(+,-,+,+,-,+) = 2e^{-2K_{1}}\cosh (2K_{3})\nonumber\\
\overline {\omega}_{36}&= & W'(+,+,+,-,-,-) = 2e^{2K_{1}}\nonumber\\
\overline {\omega}_{34}&= & W'(+,+,-,+,-,-) = 2e^{-2K_{1}}
\cosh (2K_{2}-2K_{3})\nonumber\\
\omega_{34}&= & W'(+,-,-,-,-,+) = 2e^{2K_{1}}\cosh (2K_{2})\nonumber\\
\overline {\omega}_{35}&= & W'(+,+,-,-,-,-) = 2e^{2K_{1}}
\cosh (2K_{3})\nonumber\\
\omega_{35}&= & W'(+,-,-,+,-,+) = 2e^{-2K_{1}}\nonumber\\
\overline {\omega}_{45}&= & W'(+,+,-,-,+,-) = 2e^{-2K_{1}}
\cosh (2K_{2}-2K_{3})\nonumber\\
\omega_{45}&= & W'(+,-,-,+,+,+) = 2e^{2K_{1}}\cosh (2K_{2})\nonumber\\
\overline {\omega}_{12}&= & W'(+,-,+,-,-,+) = 2e^{-2K_{1}}
\cosh (-2K_{2}+2K_{3})\nonumber\\
\omega_{12}&= & W'(+,+,+,+,-,-) = 2e^{2K_{1}}\cosh (2K_{2})\nonumber\\
\overline {\omega}_{26}&= & W'(+,+,+,-,-,+) = 2e^{2K_{1}}
\cosh (2K_{3})\nonumber\\
\omega_{26}&= & W'(+,-,+,+,-,-) = 2e^{-2K_{1}}\nonumber\\
\omega_{16}&= & W'(+,+,-,-,+,+) = 2e^{2K_{1}}\cosh (2K_{2})\nonumber\\
\overline {\omega}_{16}&= & W'(+,-,-,+,+,-) = 2e^{-2K_{1}} \cosh
(2K_{2}-2K_{3})\label{eq30}
\end{eqnarray}

Using the above expressions in Eqs. (\ref{re-eq15}), (\ref{eq16})
and (\ref{eq17}) we can obtain the expression of the free energy
of the centered honeycomb lattice Ising model.

Taking into account Eqs. (\ref{eq30}), (\ref{eq17}) and
(\ref{eq18}), the critical temperature of the model is determined
from the equation :

\begin{eqnarray}
e^{2K_{1}}+e^{-2K_{1}}\cosh (4K_{2}-2K_{3})&+& 2e^{-2K_{1}}\cosh (2K_{3})+
2e^{2K_{1}}+\nonumber \\
e^{6K_{1}}\cosh (4K_{2}+2K_{3})+e^{-6K_{1}}&= &
2\mbox{max}\{e^{2K_{1}}+
e^{-2K_{1}}\cosh (4K_{2}-2K_{3})\:\:;\:\: \nonumber\\
e^{2K_{1}}+ e^{-2K_{1}}\cosh (2K_{3})&;& e^{6K_{1}}\cosh
(4K_{1}+2K_{3})+e^{-6K_{1}}\}\label{eq31}
\end{eqnarray}
The solutions of this equation are analyzed in the next section.

We think that with the two cases studied above, the
reader will be able to apply this procedure to other
2D Ising models without crossing bonds as, for
instance, the Ising model on the centered square lattice.
After decimation of the central spin in each square, this
model can be mapped into a special case of the 16-vertex
model, by following the same procedure that we have employed
for the honeycomb lattice model.

\subsection{Disorder solutions for two-dimensional
Ising models}

Disorder solutions\index{disorder solutions} are very useful for
clarifying the phase diagrams of anisotropic models and also imply
constraints on the analytical behavior of the partition function
of these models.

A great variety of anisotropic models ( with different
coupling constants in the different directions of the
lattice ) are known to posses remarkable submanifolds in the
space of parameters, where the partition function is
computable and takes a very simple form. These are the
disorder solutions.

All the methods applied for obtaining these solutions rely
on the same mechanism : a certain local decoupling of the
degrees of freedom of the model, which results in an
effective reduction of dimensionality for the lattice
system. Such a property is provided by a simple local
condition imposed on the Boltzmann weights of the elementary
cell generating the lattice.\cite{Mail}

Some completely integrable models present disorder
solutions, e.g. the triangular Ising model and
the symmetric 8-vertex model. But very important models
that are not integrable, also present this type of
solutions, e.g. the triangular Ising model with a field, the
triangular q-state Potts model, and the general 8-vertex
model. Here we will consider only two dimensional Ising models.

\begin{figure}[th]              
\centerline{\psfig{file=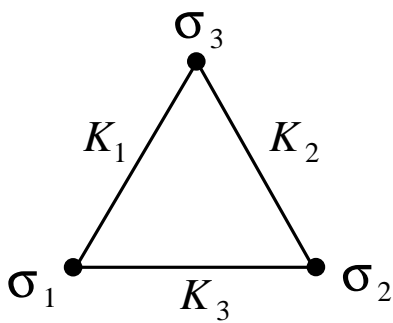,width=2.2in,angle=0}}
\vspace*{8pt} \caption{ The elementary cell of the triangular
lattice, with three interactions $K_{1}$, $K_{2}$, and $K_{3}$.
\label{re-fig15}}
\end{figure}

In order to introduce the method, we will analyze, at the first
place, the simplest case, i.e. the anisotropic Ising model on the
triangular lattice ( see Fig. \ref{re-fig15}).

The Boltzmann weight of the elementary cell is
\begin{equation}
W(\sigma_{1},\sigma_{2}, \sigma_{3})= \exp[\frac {1}{2}
(K_{1}\sigma_{1}\sigma_{3}+
K_{2}\sigma_{2}\sigma_{3}+K_{3}\sigma_{1}\sigma_{2})]\label{eq32}
\end{equation}

In every case, the local criterion will be defined by the
following condition : after summation over some of its spins
( to be defined in each case ) , the Boltzmann weight
associated with the elementary cell of the model must not
depend on the remaining spins any longer. For instance, for
the triangular lattice, we will require
\begin{equation}
\sum_{\sigma_{3}} W(\sigma_{1},\sigma_{2}, \sigma_{3})=\lambda
(K_{1},K_{2},K_{3})\label{eq33}
\end{equation}
where $\lambda$ is a function only of $K_{1}$, $K_{2}$ and $K_{3}$
( it is independent of $\sigma_{1}$ and $\sigma_{2}$). By using
Eq. (\ref{eq32}) we find
\begin{equation}
\sum_{\sigma_{3}} W(\sigma_{1},\sigma_{2}, \sigma_{3})= \exp(\frac
{1}{2} K_{3}\sigma_{1}\sigma_{2})\cosh [\frac {1}{2}(
K_{1}\sigma_{1}+K_{2}\sigma_{2})]\label{eq34}
\end{equation}
But, as it is well known, we can write
\begin{equation}
\cosh [\frac {1}{2}(K_{1}\sigma_{1}+K_{2}\sigma_{2})]= A
\exp(K\sigma_{1}\sigma_{2})\label{eq35}
\end{equation}
with
\begin{equation}
A=[\cosh (\frac {K_{1}+K_{2}}{2})\cosh (\frac
{K_{1}-K_{2}}{2})]^{\frac {1}{2}}\label{eq36}
\end{equation}
\begin{equation}
K=\frac {1}{2}\log [\frac {\cosh (\frac {K_{1}+K_{2}}{2})} {\cosh
(\frac {K_{1}-K_{2}}{2})}]\label{eq37}
\end{equation}

In order that $\sum_{\sigma_{3}} W(\sigma_{1},\sigma_{2},\sigma_{3})$
be independent of $\sigma_{1}$ and $\sigma_{2}$ we must impose
the condition $K=-\frac {1}{2}K_{3}$. From this condition we find
\begin{equation}
e^{K_{3}}\cosh (\frac {K_{1}+K_{2}}{2})= \cosh (\frac
{K_{1}-K_{2}}{2})\label{eq38}
\end{equation}
from which we can determine the expression of $\lambda$:
\begin{equation}
\lambda (K_{1},K_{2},K_{3}) =[\cosh (\frac {K_{1}+K_{2}}{2}) \cosh
(\frac {K_{1}-K_{2}}{2})]^{\frac {1}{2}}\label{eq39}
\end{equation}
It is easy to verify that  Eq.(\ref{eq38}) can be written as
\begin{equation}
\tanh (K_{1})\tanh (K_{2})+\tanh (K_{3}) =0\label{eq40}
\end{equation}

This 2D subvariety in the space of parameters
is called the disorder variety of the model.

\begin{figure}[th]              
\centerline{\psfig{file=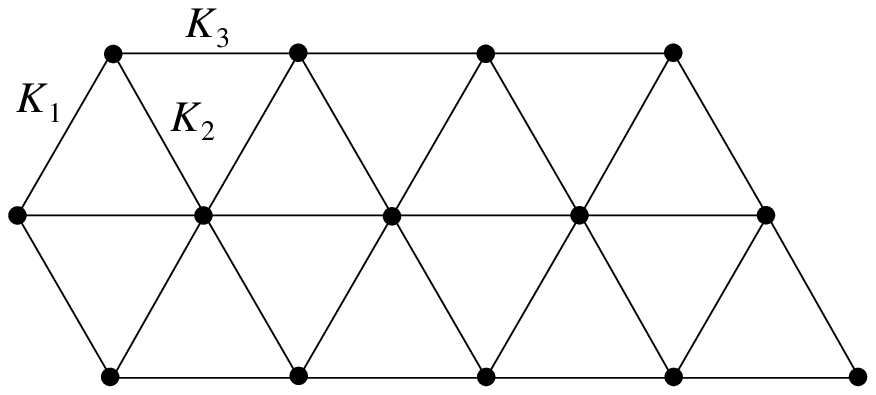,width=2.2in,angle=0}}
\vspace*{8pt} \caption{ Two layers of the triangular lattice.
\label{re-fig16}}
\end{figure}

Let us now impose particular boundary conditions for the lattice (
see Fig. \ref{re-fig16}) : on the upper layer, all interactions
are missing, so that the spins of the upper layer only interact
with those of the lower one. It immediately follows that if one
sums over all the spins of the upper layer and if one requires the
disorder condition Eq. (\ref{eq40}) , the same boundary conditions
reappear for the next layer.

Iterating the procedure leads one to an exact expression for the
partition function, when restricted to subvariety Eq.
(\ref{eq40}):
\begin{equation}
Z=\lambda (K_{1},K_{2},K_{3})^{N}\label{eq41}
\end{equation}

where $N$ is the number of sites of the lattice.
The free energy in the thermodynamic limit is given by
\begin{equation}
f=-\frac {1}{2\beta }\log [\cosh (\frac {K_{1}+K_{2}}{2}) \cosh
(\frac {K_{1}-K_{2}}{2})]\label{eq42}
\end{equation}

The partition function Eq. (\ref{eq41}) corresponds to lattices
with unusual boundary conditions. In the physical domain, where
the coupling constants are real, these do not affect the partition
function per site ( or the free energy per site) in the
thermodynamic limit, and the expression Eq. (\ref{eq42}) also
corresponds to the free energy per site with standard periodic
boundary conditions. On the contrary, in the non-physical domain (
complex coupling constants), the boundary conditions are known to
play an important role, even after  taking the thermodynamic
limit.

Let us consider now the Kagom\'{e} lattice Ising model with
two-spin interactions between nn and
nnn, studied in the next section. If we apply
the same procedure that for the triangular lattice Ising
model, we obtain for the disorder variety:
\begin{equation}
e^{4K_{2}}=\frac {2(e^{4K_{1}}+1)}{e^{8K_{1}}+3}\label{eq43}
\end{equation}

This disorder variety does not have intersection with the critical
variety of the model.

Following the method that we have exposed for the Ising model on
the triangular
lattice, the reader will be able to find the
disorder varieties for other 2D Ising models
with anisotropic interactions.

\section{Reentrance in exactly solved frustrated Ising spin
systems}

In this section, we show and discuss the phase diagrams of several
selected 2D frustrated Ising systems that have been recently
solved.  For general exact methods, the reader is referred to the
book by Baxter,\cite{Bax} and to the preceding section.  In the
following, we consider only frustrated systems that exhibit the
{\bf reentrance phenomenon}.\index{reentrance} A reentrant phase
can be defined as a phase with no long-range order, or no order at
all, occurring in a region below an ordered phase on the
temperature ($T$) scale.  A well-known example is the reentrant
phase in spin-glasses (see the review by Binder and Young\cite
{Bin}). The origin of the reentrance in spin-glasses is not well
understood. It is believed that it is due to a combination of
frustration and bond disorder. In order to see the role of the
frustration alone, we show here the exact results on a number of
periodically frustrated Ising systems.  The idea behind the works
shown in this section is to search for the ingredients responsible
for the occurrence of the reentrant phase.  Let us review in the
following a few models showing a reentrant phase.  Discussion on
the origin of the reentrance
 will be given in the conclusion.

\subsection{Centered square lattice}

\begin{figure}[th]              
\centerline{\psfig{file=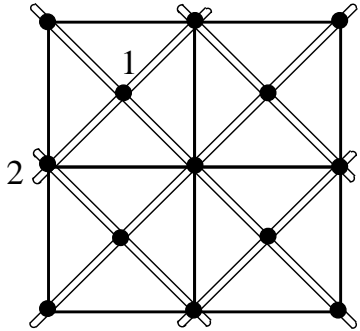,width=2.0in,angle=0}}
\vspace*{8pt} \caption{ Centered square lattice. Interactions
between nn and nnn, $J_{1}$ and $J_{2}$, are denoted by white and
black bonds, respectively. The two sublattices are numbered 1 and
2. \label{re-fig17}}
\end{figure}

Even before the concept of frustration was introduced,\cite{Tou}
systems with competing interactions were found to possess rich
critical behavior and non-trivial ordered states. Among these
models, the centered square lattice  Ising model (see Fig.
\ref{re-fig17}), introduced by Vaks et al,\cite{Vaks} with nn and
nnn interactions, $J_{1}$ and $J_{2}$, respectively, is to our
knowledge the first exactly soluble model which exhibits
successive phase transitions with a reentrant paramagnetic phase
at low $T$. Exact expression for the free energy,  some
correlation functions, and the magnetization of one sublattice
were given in the original work of Vaks et al.

We distinguish two sublattices 1 and 2. Sublattice 1 contains the
spins at the square centers, and sublattice 2 generates a square
lattice with interaction $J_{2}$ in both horizontal and vertical
directions. Spins of sublattice 1 interacts only with spins of
sublattice 2 via diagonal interactions $J_{1}$. The ground state
properties of this model are as follows : for $a = J_{2}/\mid
J_{1}\mid > - 1$, spins of sublattice 2 orders ferromagnetically
and the spins of sublattice 1 are parallel (antiparallel) to the
spins of sublattice 2 if $J_{1} > 0$ ( $< 0$ ); for $a < -1$,
spins of  sublattice 2 orders antiferromagnetically, leaving the
centered spins free to flip.

\subsubsection{Phase diagram}

The phase diagram of this model is given by Vaks et al.\cite {Vaks}
Except for $a = -1$
, there is always a finite critical temperature.

\begin{figure}[th]              
\centerline{\psfig{file=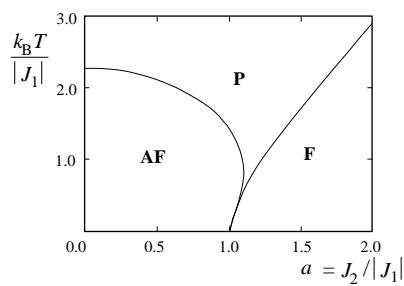,width=2.2in,angle=0}}
\vspace*{8pt} \caption{ Phase diagram of centered square lattice \cite{Vaks}.}
 \label{re-fig18}
\end{figure}

When $J_{2}$ is antiferromagnetic ($>0$) and $J_{2}/J_{1}$ is in a
small region near 1,  the system is  successively  in the
paramagnetic state, an  ordered state, the {\bf reentrant
}paramagnetic state, and another ordered state, with decreasing
temperature (see Fig. \ref{re-fig18}).

The centered square Ising lattice has been generalized to include
several kinds of interaction.\cite {Mo,Chi}  For example, when the
vertical interaction $J_{2}$ is different from the horizontal one,
say $J_{3}$, the phase diagram becomes more complicated.

\subsubsection{Nature of ordering and disorder solutions}\index{disorder
solutions}

\begin{figure}[th]              
\centerline{\psfig{file=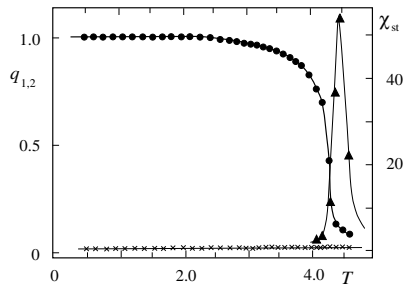,width=2.2in,angle=0}}
\vspace*{8pt} \caption{ Temperature dependence of sublattice
Edwards-Anderson order parameters, $q_{1}$ and $q_{2}$ (crosses
and black circles, respectively) in the case $a = J_{2}/\mid
J_{1}\mid = - 2$, by Monte Carlo simulation. Susceptibility
calculated by fluctuations of magnetization of sublattice 2 is
also shown. The lattice used contains $N = 2\times 60 \times 60$
spins with periodic boundary conditions.}\cite{Aza89a}
\label{re-fig19}
\end{figure}

For the sake of simplicity, let us consider hereafter the case of
nn and nnn interactions only, namely $J_{1}$ and $J_{2}$
($J_{3}=J_{2}$). Note that though an exact critical line was
obtained,\cite {Vaks} the order parameter was not calculated,
though the magnetization of one sublattice were given in the
original work of Vaks et al.\cite{Vaks} Later, Choy and Baxter
\cite {Choy} have obtained the total magnetization for this model
in the ferromagnetic region. However, the ordering in the
antiferromagnetic (frustrated) region has not been exactly
calculated, despite the fact that it may provide an interesting
ground for  understanding  the reentrance phenomenon. We have
studied this aspect by means of Monte Carlo (MC)
simulations.\cite{Aza89a} The question which naturally arises is
whether or not the disorder of sublattice 1 at $T = 0$ remains at
finite $T$. If the spins of sublattice 1 remains disordered at
finite $T$ in the antiferromagnetic region we have a remarkable
kind of ordered state: namely the coexistence between order and
disorder. This behavior has been observed in three-dimensional
Ising spin models\cite {Blan,Diep85b} and in an exactly soluble
model (the Kagom\'{e} lattice).\cite{Aza87} In the latter system,
which is similar to the present model (discussed in the next
subsection), it was shown that the coexistence of order and
disorder at finite $T$ shed some light on the reentrance
phenomena. To verify the coexistence between order and disorder in
the centered square lattice, we have performed Monte Carlo (MC)
simulations. The results for the Edwards-Anderson sublattice order
parameters $q_{i}$ and the staggered susceptibility of sublattice
2 , as functions of $T$, are shown in Fig. \ref{re-fig19} in the
case $a = - 2$.

As is seen, sublattice 2 is ordered up to the transition at
$T_{c}$ while sublattice 1 stays disordered at  all $T$. This
result shows a new example where order and disorder  coexists in
an equilibrium state.  This result supports the conjecture
formulated by Azaria et al, \cite{Aza87} namely  the coexistence
of order and disorder is a necessary condition for the reentrant
behavior to occur. The partial disorder just compensates the loss
of entropy due to the partial ordering of the high-$T$ phase. In a
previous paper,\cite{Aza87} the importance of the disorder line in
understanding the reentrance phenomenon has been emphasized. There
has been suggested that this type of line may be necessary for the
change of ordering from the high-$T$ ordered phase to the low-$T$
one. In the narrow reentrant paramagnetic region, preordering
fluctuations with different symmetries exist near each critical
line. Therefore the correlation functions  change their behavior
as the temperature is varied in the reentrant paramagnetic region.
As a consequence of the change of symmetries there exist spins for
which the two-point correlation  function (between nn spins) has
different signs, near the two critical lines , in the reentrant
paramagnetic region. Hence it is reasonable to expect that it has
to vanish at a disorder temperature $T_{D}$ . This point can be
considered as a non-critical transition point which separates two
different paramagnetic phases. The two-point correlation function
defined above may be thought of as a non-local 'disorder
parameter'. This particular point is just the one which has been
called a disorder point by Stephenson\cite {Ste} in analyzing the
behavior of correlation functions for systems with competing
interactions. For the centered square lattice Ising model
considered here, the Stephenson disorder line\index{disorder
lines} is\cite{Aza89a}
\begin{equation}
\cosh (4J_{1}/k_{B}T_{D}) = \exp(-4J_{2}/k_{B}T_{D})\label{eq44}
\end{equation}
The two-point correlation function at $T_{D}$ between spins of
sublattice 2  separated by a distance r is zero for odd r and decay
like  $r^{- 1/2}[\tanh (J_{2}/k_{B}T_{D})]^{r}$  for r even.\cite{Ste}
However,there
is {\it no dimensional reduction} on the Stephenson line given above.
Usually, one defines the disorder point as the temperature where
there is an effective reduction of dimensionality in such a way
that physical quantities become simplified spectacularly.\cite{Mail}
In general, these two types of disorder line are equivalent, as
for example , in the case of the Kagom\'{e} lattice Ising model (see below).
This is not the case here. In order to calculate this disorder line for
the centered square lattice, we recall that
this  model is equivalent to an 8-vertex
model that verifies the free-fermion condition.\cite{Wu}   The  disorder
line corresponding to dimensional reduction, was given for the
general 8-vertex model by Giacomini.\cite {Giaco86}
When this result is applied to
the centered square lattice, one finds that the disorder variety is given by
\begin{equation}
\exp(4J_{2}/k_{B}T) = (1-i\sinh (4J_{1}/k_{B}T))^{-1}\label{eq45}
\end{equation}
where $i^{2} = -1$. This disorder line lies on the unphysical
(complex) region of the parameter space of this system. When
calculated on the line Eq. (\ref{eq45}), the magnetization of this
model, evaluated recently by Choy and Baxter \cite {Choy} in the
ferromagnetic region, becomes singular, as it is usually the case
for the disorder solutions\index{disorder solutions} with
dimensional reduction.\cite {Bax86} Since in the centered square
lattice, the two kinds of disorder line are not equivalent, we
conclude, according to the arguments presented above, that the
Stephenson disorder line Eq. (\ref{eq44}) is the relevant one for
the reentrance phenomenon.

Disorder solutions have recently found interesting applications,
as for example in the problem of cellular automata (for a review
see Rujan\cite{Rujan}). Moreover, they also serve to built a new
kind  of series expansion for lattice spin systems.\cite{Mail}

\subsection{Kagom\'{e} lattice}
\subsubsection{Model with nn and nnn interactions}

Another model of interest is the Kagom\'{e}
lattice\index{Kagom\'{e} lattice} shown in Fig. \ref{re-fig7}. The
Kagom\'{e} Ising lattice with nn interaction $J_{\rm 1}$ has been
solved a long time ago\cite{Ka/Na} showing no phase transition at
finite $T$ when $J_{1}$ is antiferromagnetic. Taking into account
the nnn interaction $J_{2}$ [see Fig.\ref{re-fig7} and Eq.
(\ref{re-eq19})], we have solved\cite{Aza87} this model by
transforming it into a 16-vertex model which satisfies the
free-fermion condition.

\begin{figure}[th]              
\centerline{\psfig{file=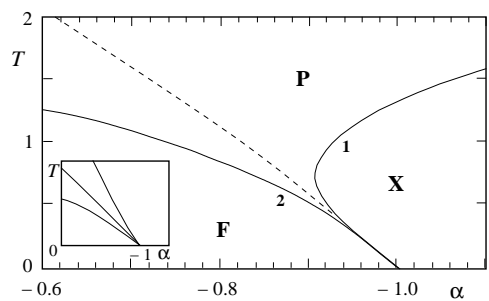,width=2.7in,angle=0}}
\vspace*{8pt} \caption{ Phase diagram of the Kagom\'{e} lattice
with nnn interaction in the region $J_{1} > 0$ of the space
($\alpha=J_{2}/J_{1}, T$). $T$ is measured in the unit of
$J_{1}/k_{B}$.  Solid lines are critical lines, dashed line is the
disorder line. P, F and X stand for paramagnetic, ferromagnetic
and partially disordered phases, respectively.  The inset shows
schematically enlarged region of the endpoint. \label{re-fig20}}
\end{figure}

The critical condition is given by Eq. (\ref{eq25}).  For the
whole phase diagram, the reader is referred to the paper by Azaria
et al.\cite{Aza87} We show in Fig. \ref{re-fig20} only the small
region of $J_{2}/J_{1}$ in the phase diagram which has the
reentrant paramagnetic phase and a disorder line.

The phase X indicates a partially ordered phase \index{partially
ordered phase} where the central spins are free (the nature of
ordering was determined by MC simulations).\cite{Aza87} Here
again, the reentrant phase takes place between a low-$T$ ordered
phase and a partially disordered phase. This suggests that a
partial disorder at the high-$T$ phase is necessary to ensure that
the entropy is larger than that of the reentrant phase.

\subsubsection{Generalized Kagom\'{e} lattice}

\begin{figure}[th]              
\centerline{\psfig{file=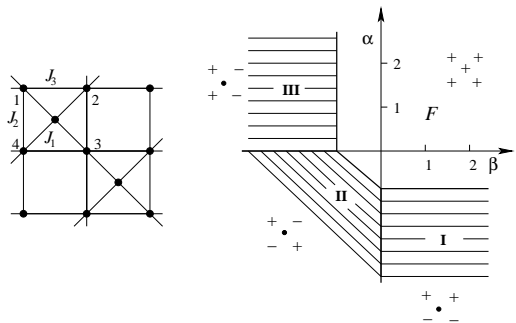,width=2.7in,angle=0}}
\vspace*{8pt} \caption{ Left: Generalized Kagom\'{e} lattice:
diagonal, vertical and horizontal  bonds denote the interactions
$J_{1}$, $J_{2}$ and $J_{3}$, respectively. Right: Phase diagram
of the ground state shown in the plane ($\alpha =J_{2}/J_{1},
\beta = J_{3}/J_{1}$). Heavy lines separate different phases and
spin configuration of each phase is indicated (up, down and free
spins are denoted by +, - and o, respectively).  The three kinds
of partially disordered phases and the ferromagnetic phase are
denoted by I, II , III and F, respectively. \label{re-fig21}}
\end{figure}

When all the interactions are different in the model shown in Fig.
\ref{re-fig7}, i.e. the horizontal bonds $J_{3}$, the vertical
bonds $J_{2}$ and the diagonal ones are not equal (see Fig.
\ref{re-fig21}), the phase diagram becomes complicated with new
features:\cite {Diep91b} in particular, we show that the
reentrance can occur in an {\it infinite region} of phase space.
In addition, there may be {\it several reentrant phases} occurring
for a given set of interactions when $T$ varies.

The Hamiltonian is written as
\begin{equation}
H=-J_{1}\sum_{(ij)} \sigma_{i}\sigma_{j}-J_{2}\sum_{(ij)}
\sigma_{i}\sigma_{j}-J_{3}
\sum_{(ij)}\sigma_{i}\sigma_{j}\label{eq46}
\end{equation}
where $\sigma_{i}=\pm 1$  is an Ising spin occupying the lattice site i ,
and the
first, second, and third sums run over the spin pairs
connected by
diagonal, vertical and horizontal bonds, respectively.
When
$J_{2}=0$ and $J_{1}=J_{3}$, one recovers the original nn Kagom\'{e}
lattice.\cite {Ka/Na}  The effect of $J_{2}$ in the case $J_{1}=J_{3}$
has been shown above.


The phase diagram at temperature $T=0$ is shown in Fig.
\ref{re-fig21} in the space ($\alpha = J_{2}/J_{1}$, $\beta =
J_{3}/J_{1}$) for positive $J_{1}$. The ground- state  spin
configurations are also displayed. The hatched regions indicate
the three partially disordered phases (I, II, and III) where the
central spins are free.  Note that the phase diagram is
mirror-symmetric with respect to the change of the sign of
$J_{1}$. With negative $J_{1}$ , it suffices to reverse the
central spin in the spin configuration shown in Fig.
\ref{re-fig21}. Furthermore, the interchange of $J_{2}$ and
$J_{3}$ leaves the system invariant, since it is equivalent to a
$\pi /2$ rotation of the lattice. Let us consider the effect of
the temperature on the phase diagram shown in Fig. \ref{re-fig21}.
Partial disorder in the ground state often gives rise to the
reentrance phenomenon as in systems shown above.  Therefore,
similar effects are to be expected in the present system.  As it
will be shown below, we find a new and richer behavior of the
phase diagram:  in particular,  the reentrance region is found to
be extended to infinity, unlike systems previously studied, and
for some given set of interactions, there exist {\it two disorder
lines} \index{disorder lines} which divide the paramagnetic phase
into regions of different kinds of fluctuations with a reentrant
behavior.

Following the method exposed in section 3, one obtains a
checkerboard Ising model with multispin interactions. This
resulting model is equivalent to a symmetric 16-vertex model which
satisfies the free-fermion condition.\cite {Gaff,Suzu,Wu72} The
critical temperature of the model is given by
\begin{equation}
\cosh (4K_{1}) \exp (2K_{2}+2K_{3}) +\exp (-2K_{2}-2K_{3}) =
2\cosh (2K_{3} - 2K_{2}) \pm 4\cosh (2K_{1})\label{eq47}
\end{equation}
Note that Eq. (\ref{eq47}) is invariant when changing
$K_{1}\rightarrow -K_{1}$  and interchanging $K_{2}$ and $K_{3}$
as stated earlier. The phase diagram in the three-dimensional
space ($K_{1},K_{2},K_{3}$) is rather complicated to show.
Instead, we  show in the following the phase diagram in the plane
($\beta =J_{3}/J_{1},T$) for typical values of $\alpha
=J_{2}/J_{1}$. To describe each case and to follow the evolution
of the phase diagram, let us go in the direction of decreasing
$\alpha$ :

\begin{center}
A.   $\alpha>0$
\end{center}

\begin{figure}[th]              
\centerline{\psfig{file=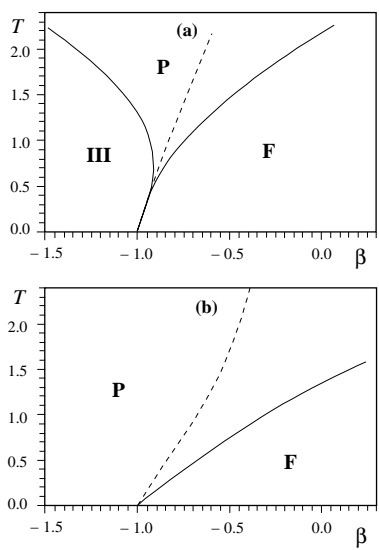,width=2.0in,angle=0}}
\vspace*{8pt} \caption{ Phase diagram in the plane ($\beta
=J_{3}/J_{1},T$) for positive values of $\alpha =J_{2}/J_{1}$:(a)
$\alpha =1$, (b) $\alpha=0$. Solid lines are critical lines which
separate different phases: paramagnetic (P), ferromagnetic (F),
partially disordered phase of type III (III). Dotted line shows
the disorder line. \label{re-fig22}}
\end{figure}

This case is shown in Fig. \ref{re-fig22}. Two critical lines are
found with a paramagnetic reentrance having a usual shape (Fig.
\ref{re-fig22}a) between the partially disordered (PD) phase of
type III (see Fig. \ref{re-fig21}) and the ferromagnetic (F) phase
with an endpoint at $\beta =-1$. The width of the reentrance
region $[-1, \beta_{1}]$ decreases with decreasing $\alpha$ , from
$\beta_{1}=0$ for $\alpha$ at infinity to $\beta_{1}=-1$ for
$\alpha=0$ (zero width).    Note that as $\alpha$ decreases, the
PD phase III is depressed and disappears at $\alpha=0$, leaving
only the F phase (one critical line, see Fig. \ref{re-fig22}b).
The absence of order at zero $\alpha$ for $\beta$ smaller than -1
results from the fact that in the ground state, this region of
parameters corresponds to a superdegenerate line separating the
two PD phases II and III (see Fig. \ref{re-fig21}).  So, along
this line, the disorder contaminates the system for all $T$.  As
for disorder solutions,\index{disorder solutions} for positive
$\alpha$ we find in the reentrant paramagnetic region a disorder
line with dimension reduction\cite {Ste} given by
\begin{equation}
\exp(4K_{3}) = 2\cosh
(2K_{2})/[\cosh(4K_{1})\exp(2K_{2})+\exp(-2K_{2})]\label{eq48}
\end{equation}
This is shown by the dotted lines in Fig. \ref{re-fig22}.

\begin{center}
B. $0 > \alpha > -1$
\end{center}

In this range of $\alpha$, there are three critical lines. The
critical line separating the F and P phases and the one separating
the PD phase I from the P phase have a common horizontal asymptote
as $\beta$ tends to infinity .  They form a reentrant paramagnetic
phase between the F phase and the PD phase I for positive b
between a value $\beta_{2}$ and infinite $\beta$ (Fig. 23).
Infinite region of reentrance like this has never been found
before this model.  As $\alpha$ decreases, $\beta_{2}$ tends to
zero and the F phase is contracted.  For $\alpha<-1$, the F phase
disappears together with the reentrance.

\begin{figure}[th]              
\centerline{\psfig{file=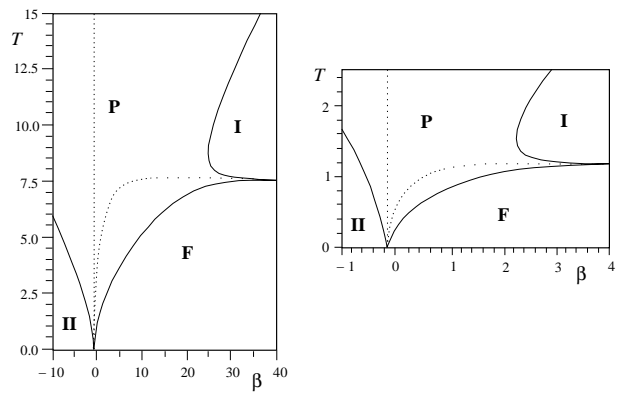,width=3.2in,angle=0}}
\vspace*{8pt} \caption{ Phase diagram in the plane ($\beta
=J_{3}/J_{1},T$) for negative values of $\alpha =J_{2}/J_{1}$.
Left: $\alpha =-0.25$, Right: $\alpha=-0.8$. Solid lines are
critical lines which separate different phases: paramagnetic (P),
ferromagnetic (F), partially disordered phases of type I and III.
Dotted lines show the disorder lines. \label{re-fig23}}
\end{figure}

\begin{figure}[th]              
\centerline{\psfig{file=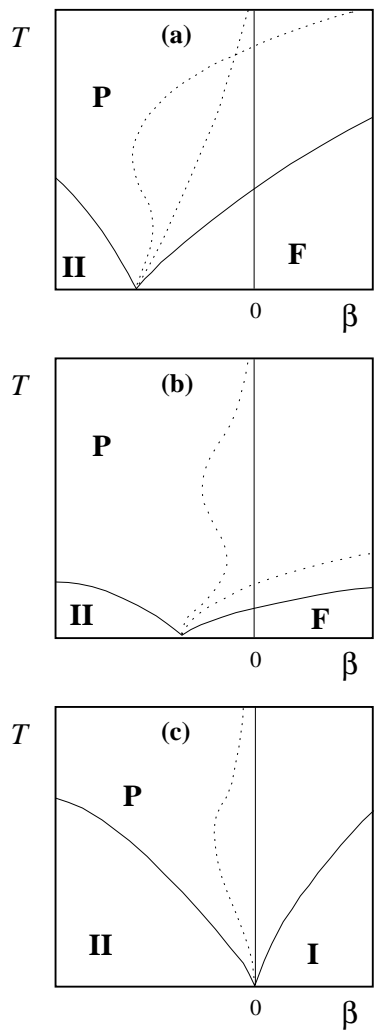,width=2.0in,angle=0}}
\vspace*{8pt} \caption{ The behavior of the disorder lines
(dotted) is schematically enlarged in the case (a) $\alpha
=-0.25$, (b) $\alpha =-0.8$, (c) $\alpha=-1.5$. \label{re-fig24}}
\end{figure}

In the interval $0 > \alpha > -1$, the phase diagram possesses two
disorder lines,\index{disorder lines} the first being given by Eq.
(\ref{eq48}), and the second  by
\begin{equation}
\exp(4K_{3}) = 2\sinh
(2K_{2})/[-\cosh(4K_{1})\exp(2K_{2})+\exp(-2K_{2})]\label{eq49}
\end{equation}
These two disorder lines are issued from a point near $\beta =-1$
for small negative $\alpha$;  this point tends to zero as $\alpha$
tends to -1. The  disorder line given by Eq. (\ref{eq49}) enters
the reentrant region which separates the F phase and the PD phase
I (Fig. \ref{re-fig23}, left), and the one given by Eq.
(\ref{eq48}) tends to infinity with the asymptote $\beta =0$ as $T
\rightarrow \infty$. The most striking feature is the behavior of
these two disorder lines at low $T$:  they cross each other in the
P phase for  $0 > \alpha > -0.5$, forming regions of fluctuations
of different nature (Fig. \ref{re-fig24}a).   For $-0.5
> \alpha > -1$, the two disorder lines do no longer cross each
other (see Fig. \ref{re-fig24}b). The one given by Eq.
(\ref{eq48}) has a reentrant aspect: in a small region of negative
values of $\beta$, one crosses three times this line in the P
phase with decreasing $T$.

\begin{center}
C.  $\alpha \leq -1$
\end{center}

\begin{figure}[th]              
\centerline{\psfig{file=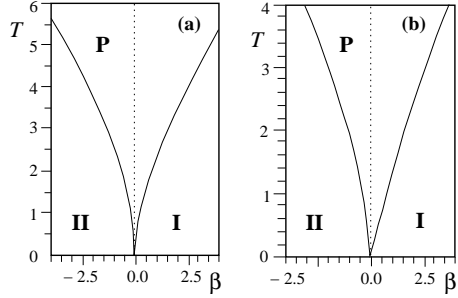,width=2.4in,angle=0}}
\vspace*{8pt} \caption{ The same caption as that of Fig.
\ref{re-fig23} with (a) $\alpha =-1$, (b) $\alpha=-1.5$.
\label{re-fig25}}
\end{figure}

For  $\alpha$ smaller than -1, there are two critical lines and no
reentrance (Fig. \ref{re-fig25}). Only the disorder line given by
Eq. (\ref{eq48}) survives with a reentrant aspect: in a small
region of negative values of $\beta$, one crosses twice this line
in the P phase with decreasing $T$.  This behavior, being
undistiguishable in the scale of Fig. \ref{re-fig25},  is
schematically enlarged in Fig. \ref{re-fig24}c. The multicritical
point where the P, I and II phases meet is found at $\beta=0$ and
$T=0$.

At this stage, it is interesting to note that while reentrance and
disorder lines occur along the horizontal axis $\alpha=-1$ and
along the vertical axis $\beta=-1$ of Fig. \ref{re-fig21} when the
temperature is switched on, the most frustrated region ($\alpha<0$
and $\beta<0$) of the ground state does not show successive phase
transitions (see Fig. \ref{re-fig25}, for example).  Therefore,
the existence of a reentrance may require a sufficient
frustration, but not overfrustration.  Otherwise, the system may
have either a PD phase (Fig. \ref{re-fig25})  or no order at all
(Fig. \ref{re-fig22}b).

The origin of the reentrance
phenomenon will be discussed again in the conclusion.

\subsection{Centered honeycomb lattice}

In order to find common aspects of the reentrance phenomenon, we
have constructed a few other models which possess a partially
disordered phase next to an ordered phase in the ground state. Let
us mention here the anisotropic centered honeycomb lattice shown
in Fig. \ref{re-fig11}.\cite {Diep91a} The Hamiltonian is given by
Eq. (\ref{eq25}), with three kinds of interactions $J_{1}$,
$J_{2}$, and $J_{3}$ denoting the interactions between the spin
pairs connected by heavy, light, and double-light bonds,
respectively. We recall that when $J_{2}=J_{3}=0$, one recovers
the honeycomb lattice, and when $J_{1}=J_{2}=J_{3}$ one has the
triangular lattice.

\begin{figure}[th]              
\centerline{\psfig{file=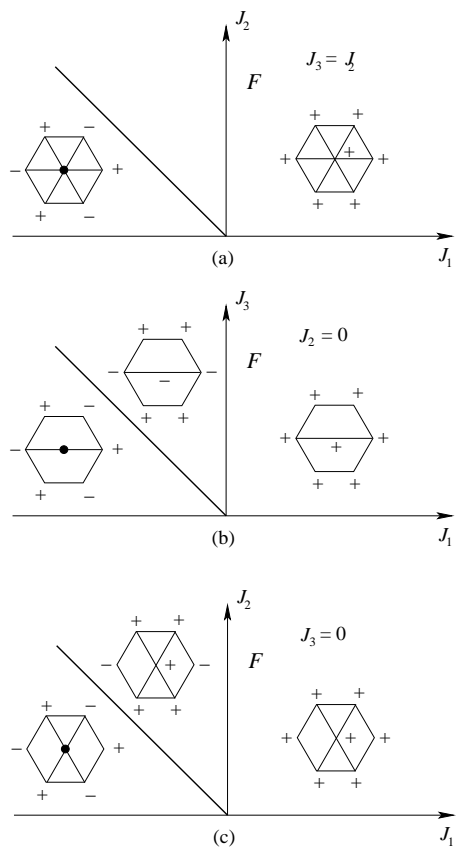,width=2.4in,angle=0}}
\vspace*{8pt} \caption{ Phase diagram of the ground state shown in
the space: (a) ($J_{1},J_{2} = J_{3}$) ;  (b)  ($J_{1},J_{3}$)
with $J_{2}=0$) ; (c) ($J_{1},J_{2}$) with $J_{3}=0$). Heavy lines
separate different phases and spin configuration of each phase is
indicated (up, down and free spins are denoted by +, - and o,
respectively). \label{re-fig26}}
\end{figure}

Fig. \ref{re-fig26} shows the phase diagram at  temperature $T=0$
for three cases ($J_{1} \neq J_{2} = J_{3}$), ( $J_{1} \neq J_{3}
, J_{2}=0$) and ($J_{1} \neq J_{2}, J_{3}=0$). The ground-state
spin configurations are also indicated.

The phase diagram is symmetric with respect to the horizontal
axis: the transformation $(J_{2} , \sigma) \rightarrow  (-J_{2} ,
-\sigma)$ , or $(J_{3} , \sigma) \rightarrow (-J_{3} , -\sigma)$ ,
leaves the system invariant.  In each case, there is a phase where
the central spins are free to flip ("partially disordered phase").
In view of this common feature with other models studied so far,
one expects a reentrant phase occurring between the partially
disordered phase and its neighboring phase at finite $T$. As it
will be shown below, though a partial disorder exists in the
ground state, it does not in all cases studied here yield a
reentrant phase at finite $T$.  Only the case  ($J_{1} \neq J_{2},
J_{3}=0$) does show  a reentrance.

To obtain the exact solution of our model, we decimate the central
spin of each elementary cell of the lattice as outlined in the
section 3. The resulting model is equivalent to a special case of
the 32-vertex model\cite{Sacco} on a triangular lattice that
satisfies the free-fermion condition. The explicit expression of
the free energy as a function of interaction parameters $K_{1}$,
$K_{2}$, and $K_{3}$ is very complicated, as seen by replacing Eq.
(\ref{eq29}) in Eqs. (\ref{re-eq15}), (\ref{eq16}) and
(\ref{eq17}). The critical temperature is given by Eq.
(\ref{eq30}).

We have analyzed, in particular,  the three cases
($K_{1} \neq K_{2} = K_{3}$) ,  ($K_{1} \neq K_{3} , K_{2}=0$) and
($K_{1} \neq K_{2}, K_{3}=0$).

When $K_{2}=K_{3}$, the critical line obtained from
Eq.(\ref{eq30}) is
\begin{equation}
\exp(3K_{1}) \cosh (6K_{2}) + \exp(-3K_{1}) = 3[\exp(K_{1}) +
\exp(-K_{1}) \cosh (2K_{2})]\label{eq50}
\end{equation}

In the case $K_{2}=0$, the critical line is given by
\begin{equation}
\exp(3K_{1})\cosh (2K_{3}) + \exp(-3K_{1}) = 3[\exp(K_{1}) +
\exp(-K_{1})\cosh (2K_{3})]\label{eq51}
\end{equation}

Note that these equations are invariant with respect to the
transformation $K_{2} \rightarrow -K_{2}$ (see Eq.(\ref{eq50}))
and $K_{3} \rightarrow -K_{3}$ (see Eq.(\ref{eq51})).

\begin{figure}[th]              
\centerline{\psfig{file=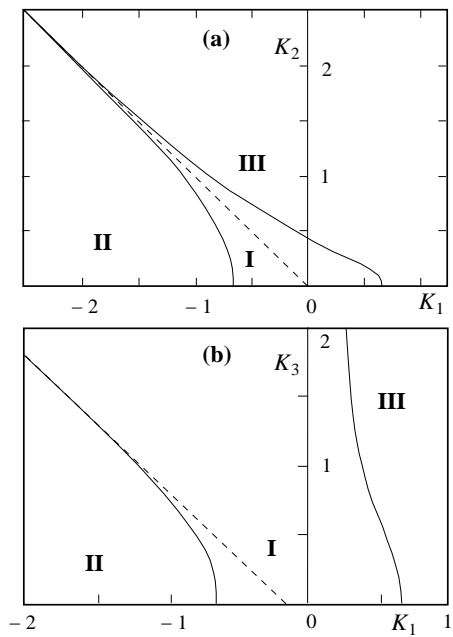,width=2.4in,angle=0}}
\vspace*{8pt} \caption{ Phase diagram shown in the space (a)
($K_{1},K_{2} = K_{3}$) ;  (b)  ($K_{1},K_{3},K_{2}=0$).  Solid
lines are critical lines which separate paramagnetic (I) ,
partially disordered (II) and ordered (III) phases.  Discontinued
lines of slope -1 are the asymptotes. \label{re-fig27}}
\end{figure}

The phase diagrams obtained from Eqs. (44) and (45) are shown in
Fig. \ref{re-fig27}a and Fig. \ref{re-fig27}b, respectively.

These two cases  do not present the reentrance phenomenon though a
partially disordered phase exists next to an ordered phase in the
ground state (this is seen by plotting a line from the origin,
i.e. from infinite $T$: this line never crosses twice a critical
line whatever its slope, i.e. the ratio $K_{2,3}/K_{1}$, is). In
the ordered phase II, the partial disorder,\index{partial
disorder} which exists in the ground state, remains so up to the
phase transition. This has been verified by examining the
Edwards-Anderson order parameter associated with the central spins
in MC simulations.\cite{Diep91a} Note that when $K_{2}= K_{3}=0$,
one recovers the transition at finite  temperature found for the
honeycomb lattice.\cite{Onsa} and when $K_{2}= K_{3}=K_{1}=-1$ one
recovers the antiferromagnetic triangular lattice which has no
phase transition at finite temperature.\cite{Wan} The case
$K_{2}=0$ (Fig. \ref{re-fig27}b) does not have a phase transition
at finite $T$ in the range $-\infty < K_{3}/K_{1} <-1$, and phase
II has a partial disorder as that in Fig. \ref{re-fig22}a.

The case $K_{3}=0$ shows on the other hand a reentrant phase.
The critical lines are
determined from the equations
\begin{equation}
\cosh (4 K_{2})=\frac {\exp(4 K_{1})+2\exp(2 K_{1})+1}{[1- \exp(4
K_{1})]\exp(2 K_{1})}\label{eq52}
\end{equation}
\begin{equation}
\cosh (4 K_{2})=\frac {3\exp(4 K_{1})+2\exp(2 K_{1})-1}{[ \exp(4
K_{1})-1]\exp(2 K_{1})}\label{eq53}
\end{equation}

\begin{figure}[th]              
\centerline{\psfig{file=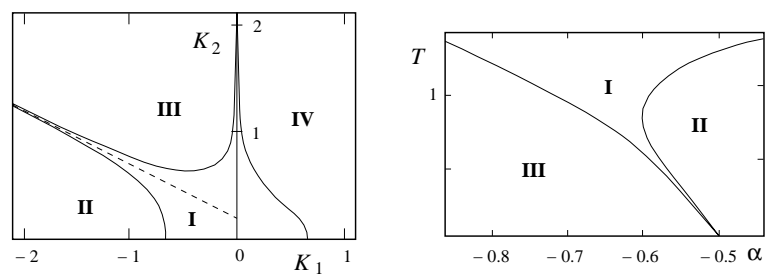,width=4.0in,angle=0}}
\vspace*{8pt} \caption{ Phase diagram of the centered honeycomb
lattice with reentrance in the space ($K_{1},K_{2}$) (left) and in
the space ($T,\alpha=K_{2}/ K_{1}$) (right). I, II, III phases are
paramagnetic, partially disordered and ordered phases,
respectively.  Discontinued line is the asymptote.
\label{re-fig28}}
\end{figure}

Fig. \ref{re-fig28} shows the phase diagram obtained from Eqs.
(\ref{eq52}) and (\ref{eq53}) .  The reentrant paramagnetic phase
goes down to zero temperature with an end point at $\alpha=-0.5$
(see Fig. \ref{re-fig28} right).

Note that the honeycomb model that we have studied here does not
present a disorder solution\index{disorder solutions} with a
dimensional reduction.

\subsection{Periodically dilute centered square lattices}

In this subsection, we show the exact results on
several periodically dilute centered square
Ising lattices by transforming them into 8-vertex models
of {\it different vertex statistical weights} that satisfy the
free-fermion condition. The dilution is introduced by taking away a
number of centered spins in a periodic manner. For a
given set of interactions
, there may be five transitions with decreasing temperature with
two reentrant paramagnetic phases.   These two phases extend to
infinity in the space of interaction parameters. Moreover,  two
additional reentrant phases are found, each in a limited region
of phase space.\cite {Diep92}

\begin{figure}[th]              
\centerline{\psfig{file=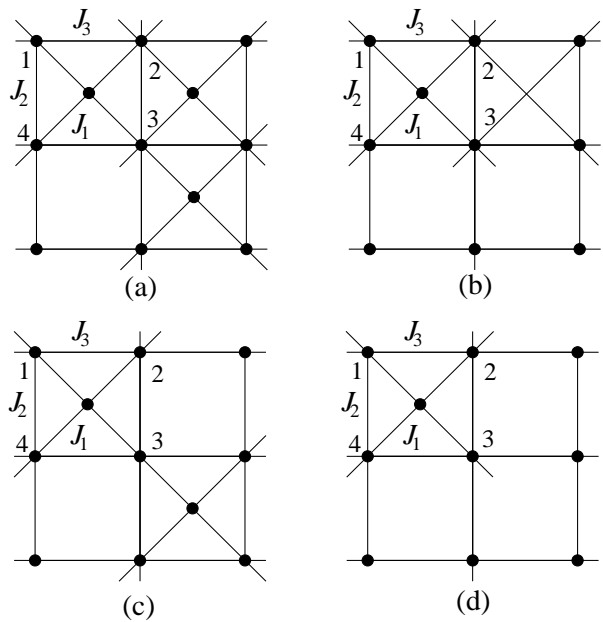,width=3.2in,angle=0}}
\vspace*{8pt} \caption{ Elementary cells of periodically dilute
centered square lattice: (a)  three-center case, (b)
two-adjacent-center case, (c) two-diagonal-center case, (d)
one-center case.  Interactions along diagonal, vertical and
horizontal bonds are $J_{1}$, $J_{2}$, and $J_{3}$, respectively.
\label{re-fig29}}
\end{figure}

Let us consider several {\bf periodically  dilute centered square
lattices}  defined from the centered square lattice shown in Fig.
\ref{re-fig29}.

The Hamiltonian of these models  is given by
\begin{equation}
H=-J_{1}\sum_{(ij)} \sigma_{i}\sigma_{j}-J_{2}\sum_{(ij)}
\sigma_{i}\sigma_{j}-J_{3}
\sum_{(ij)}\sigma_{i}\sigma_{j}\label{eq54}
\end{equation}
where $\sigma_{i}=\pm 1$  is an Ising spin occupying the lattice
site i , and the first, second and third sums run over the spin
pairs connected by diagonal, vertical and horizontal bonds,
respectively.  All these models have at least one partially
disordered phase in the ground state, caused by the competing
interactions.

The model shown in Fig. \ref{re-fig29}c is in fact the generalized
Kagom\'{e} lattice\cite {Diep91b} which is shown above. The other
models are less symmetric, require different vertex weights as
seen below.

Let us show in Fig. \ref{re-fig30} the phase diagrams, at $T=0$,
of the models shown in Figs. 25a, 25b and 25d, in the space ( $a,
b$ ) where $a =J_{2} /J_{1}$ and $b =J_{3}/J_{1}$. The spin
configurations in different phases are also displayed.  The
three-center case (Fig. \ref{re-fig30}a), has six phases (numbered
from I to VI), five of which (I, II, IV, V and VI) are partially
disordered (with, at least, one centered spin being free), while
the two-center case (Fig. \ref{re-fig30}b) has five phases, three
of which (I, IV, and V) are partially disordered.    Finally, the
one-center case has seven phases with three partially disordered
ones (I, VI and VII).
\begin{figure}[th]              
\centerline{\psfig{file=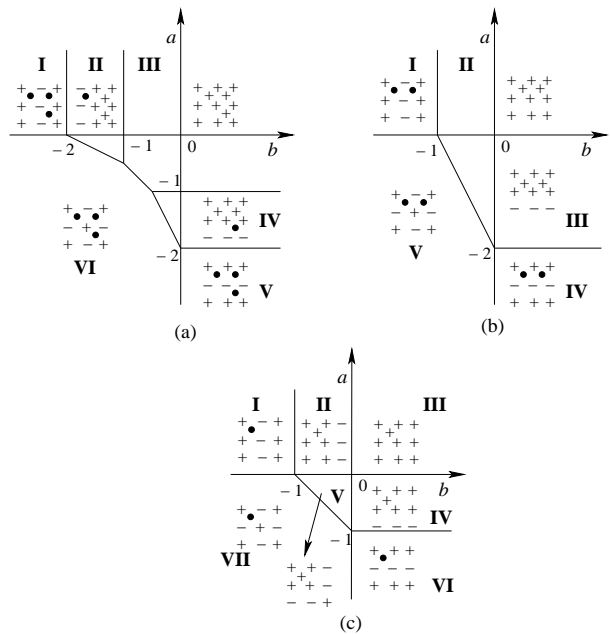,width=3.2in,angle=0}}
\vspace*{8pt} \caption{ Phase diagrams in the plane ($a =J_{2}
/J_{1}$, $b =J_{3}/J_{1}$) at $T=0$ are shown for the three-center
case (a), two-adjacent center case (b), and one-center case (c).
Critical lines are drawn by heavy lines. Each phase is numbered
and the spin configuration is  indicated (+, -, and o are up,
down, and free spins, respectively). Degenerate configurations are
obtained by reversing all spins. \label{re-fig30}}
\end{figure}

As will be shown later, in each model,  the reentrance occurs
along most of the critical lines when the temperature is switched
on. This is a very special feature of the models shown in Fig.
\ref{re-fig29} which has not been found in other models.

The partition function is written as
\begin{equation}
Z =\prod_{j} \sum_{\sigma}  W_{j}\label{eq55}
\end{equation}
where the sum is performed over all spin configurations and the product
over all elementary squares. $W_{j}$ is the statistical weight of the j-th
square. Let us denote the centered spin (when it exists) by $\sigma$ and the
spins at the square corners by $\sigma_{1}$, $\sigma_{2}$,
$\sigma_{3}$ and $\sigma_{4}$.
If the centered site exists, the statistical weight $W_{j}$ of the
square is
\begin{equation}
W_{j}= \exp[K_{1}(\sigma_{1}\sigma_{2} + \sigma_{3}\sigma_{4})
+K_{2}(\sigma_{1}\sigma_{4} + \sigma_{2}\sigma_{3})
+K_{3}\sigma(\sigma_{1}+\sigma_{2} + \sigma_{3}+\sigma_{4})]
\label{eq56}
\end{equation}
Otherwise, it is given by
\begin{equation}
W_{j}= \exp[K_{1}(\sigma_{1}\sigma_{2} + \sigma_{3}\sigma_{4})
+K_{2}(\sigma_{1}\sigma_{4} + \sigma_{2}\sigma_{3})]\label{eq57}
\end{equation}
where $K_{i} = J_{i}/ k_{B}T  (i=1,2,3)$.

In order to obtain the exact solution of these models, we decimate
the central spins of the centered squares. The resulting system is
equivalent to an eight-vertex model on a square lattice, but with
{\it different vertex weights}. For example, when one center is
missing (Fig. \ref{re-fig29}d), three squares over four have the
same weight $W_{i}$, and the fourth has  a weight  $W'_{i}\neq
W_{i}$. So, we have to define four different sublattices with
different statistical weights. The problem has been studied by
Hsue, Lin and Wu for two different sublattices\cite{Hsue} and Lin
and Wang \cite{Lin} for four sublattices. They showed that exact
solution can be obtained provided that all different statistical
weights satisfy the free-fermion condition.\cite
{Bax,Gaff,Hsue,Lin} This is indeed our case and we get the exact
partition function in terms of interaction parameters. The
critical surfaces of our models are obtained by
\begin{equation}
\Omega_{1}+\Omega_{2}+\Omega_{3}+\Omega_{4}= 2\mbox{max}
(\Omega_{1},\Omega_{2},\Omega_{3},\Omega_{4})\label{eq58}
\end{equation}
where $W_{i}$ are functions of $K_{1}$, $K_{2}$ and $K_{3}$. We explicit this equation
and we obtain a second order equation for X which is a function of
$K_{2}$ only :
\begin{equation}
A ( K_{1}, K_{3} ) X^{2} + B ( K_{1}, K_{3}) X + C ( K_{1}, K_{3})
= 0 \label{eq59}
\end{equation}
with a priori four possible values of $A$, $B$ and $C$ for each model.

For given values of $K_{1}$ and $K_{3}$, the critical surface is
determined by the value of $K_{2}$  which satisfies
Eq.(\ref{eq59}) through $X$.  $X$ must be real positive. We show
in the following the expressions of $A$, $B$ and $C$ for which
this condition is fulfilled for each model. Eq. (\ref{eq58}) may
have as much as five solutions for the critical temperature,\cite
{Lin} and the system may, for some given values of interaction
parameters, exhibit up to five phase transitions. This happens for
the model with three centers, when one of the interaction is large
positive and the other slightly negative, the diagonal one being
taken as unit. In general, we obtain one or three solutions for
$T_{c}$.

\subsubsection{Model with three centers} (Fig. \ref{re-fig29}a)

The quantities which satisfy Eq. (\ref{eq59}) are given by
\begin{eqnarray}
X&= &\exp(4K_{2}) \nonumber \\
A&= &\exp(4K_{1}) \cosh^{3}(4K_{3}) + \exp(-4K_{1})
- \cosh^{2}(4K_{3}) - \cosh(4K_{3}) \nonumber \\
B&= &\pm \{1 + 3\cosh (4K_{3}) + 8\cosh^{3}(2K_{3}) + [\cosh(4K_{3})
+ \cosh^{2}(4K_{3})] \exp(4K_{1}) \nonumber \\
 & &+ 2\exp(-4 K_{1})\} \nonumber \\
C&= &[\exp(2K_{1}) - \exp(-2K_{1})]^{2} \nonumber \\
A&= &\exp(4K_{1}) \cosh^{3}(4K_{3}) + \exp(-4K_{1})
+ \cosh^{2}(4K_{3}) + \cosh(4K_{3}) \nonumber \\
B&= &[1 + 3\cosh(4K_{3}) + 8\cosh^{3}(2K_{3})
- (\cosh(4K_{3}) + \cosh^{2}(4K_{3})) \exp(4K_{1}) \nonumber \\
 & &- 2\exp(-4K_{1})] \nonumber \\
C&= &[\exp(2K_{1}) + \exp(-2K_{1})]^{2}\label{eq60}
\end{eqnarray}

Let us describe now in detail the phase diagram of the
three-center model (Fig. \ref{re-fig29}a).

\begin{figure}[th]              
\centerline{\psfig{file=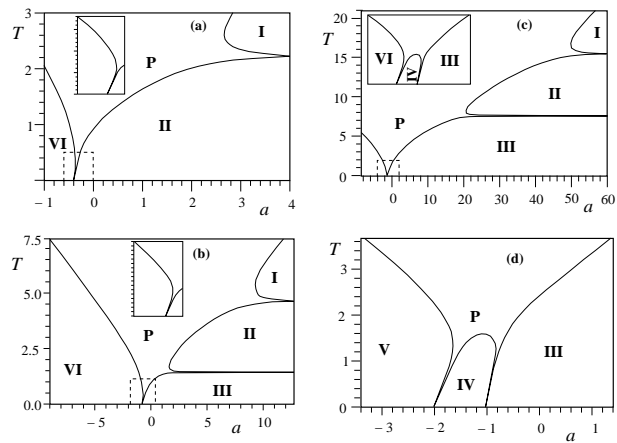,width=3.2in,angle=0}}
\vspace*{8pt} \caption{ Phase diagrams in the plane ($T, a =J_{2}
/J_{1}$) for several values of $b =J_{3}/J_{1}$:(a) $b=-1.25$, (b)
$b=-0.75$, (c) $b=-0.25$, (d) $b=0.75$.  Reentrant regions on
negative sides of $a$ (limited by discontinued lines) are
schematically enlarged in the insets.  The nature of ordering in
each phase is indicated by a number which is referred to the
corresponding spin configuration in Fig. \ref{re-fig30}. P is
paramagnetic phase. \label{re-fig31}}
\end{figure}

For clarity, we show in Fig. \ref{re-fig31} the phase diagram in
the space ( $a =J_{2} /J_{1}$, $T$) for typical values of $b
=J_{3}/J_{1}$.

For $b < -1$, there are two reentrances.  Fig. \ref{re-fig31}a
shows the case of $b = -1.25$ where the nature of the ordering in
each phase is indicated using the same numbers of corresponding
ground state configurations (see Fig. \ref{re-fig30}). Note that
all phases (I, II and VI) are partially disordered:  the centered
spins which are disordered at $T=0$ (Fig. \ref{re-fig30}a) remain
so at all $T$. As seen, one paramagnetic reentrance is found in a
small region of negative $a$ (schematically enlarged in the inset
of Fig. \ref{re-fig31}a), and the other on the positive $a$
extending to infinity.  The two critical lines in this region have
a common horizontal asymptote.

For $-1 < b < - 0.5$, there are three reentrant paramagnetic
regions as shown in Fig. \ref{re-fig31}b:  the reentrant region on
the negative $a$ is very narrow (inset), and the two on the
positive $a$ become so narrower while $a$ goes to infinity that
they cannot be seen on the scale of Fig. \ref{re-fig31}. Note that
the critical lines in these regions have horizontal asymptotes.
For a large value of $a$, one has five transitions with decreasing
$T$: paramagnetic state - partially disordered phase I - reentrant
paramagnetic phase - partially disordered phase II - reentrant
paramagnetic phase- ferromagnetic phase (see Fig. \ref{re-fig31}b
). So far, this is the first model that exhibits such successive
phase transitions with two reentrances.

For $- 0.5 < b < 0$, there is an additional reentrance for $a <
-1$:  this is shown in the inset of Fig. \ref{re-fig31}c. As $b$
increases from negative values, the ferromagnetic region (III) in
the phase diagram "pushes" the two partially disordered phases (I
and II) toward higher $T$.  At $b = 0$, these two  phases
disappear at infinite $T$, leaving only the ferromagnetic phase.
For positive b, there are thus only two reentrances remaining on a
negative region of $a$, with endpoints at $a = -2$ and $a = -1$,
at $T = 0$ (see Fig. \ref{re-fig31}d).

\subsubsection{Model with two adjacent centers} (Fig. \ref{re-fig29}b)

The quantities which satisfy Eq. (\ref{eq59}) are given by
\begin{eqnarray}
X&= &\exp(2K_{2}) \nonumber \\
A&= &\exp(2K_{1})\cosh(4K_{3}) + \exp(-2K_{1}) \nonumber \\
B&= &2[\exp(2K_{1})\cosh^{2}(2K_{3}) - \exp(-2K_{1})] \nonumber \\
C&= &\exp(2K_{1}) + \exp(-2K_{1}) \nonumber \\
A&= &\exp(2K_{1})\cosh(4K_{3}) - \exp(-2K_{1}) \nonumber \\
B&= &\pm 2[\exp(2K_{1})\cosh^{2}(2K_{3}) + \exp(-2K_{1})] \nonumber \\
C&= &\exp(2K_{1}) - \exp(-2K_{1})\label{eq61}
\end{eqnarray}

\begin{figure}[th]              
\centerline{\psfig{file=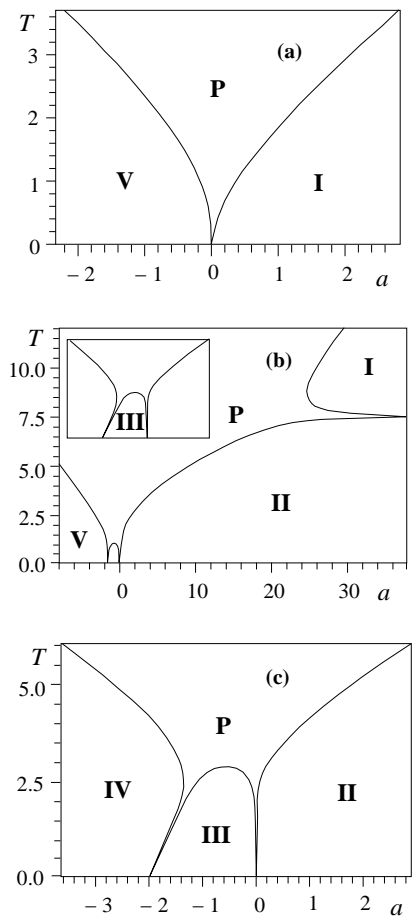,width=2.2in,angle=0}}
\vspace*{8pt} \caption{ Two-center model: the same caption as that
of Fig. \ref{re-fig31} with (a) $b=-1.25$, (b) $b=-0.25$, (c)
$b=2$. \label{re-fig32}}
\end{figure}

The phase diagram is shown in Fig. \ref{re-fig32}.

For $b < -1$, this model shows only one transition at a finite $T$
for a given value of $a$, except when $a = 0$ where the
paramagnetic state goes down to $T=0$( see Fig. \ref{re-fig32}a).

However, for $-1 < b < 0$, two reentrances appear, the first one
separating phases I and II goes to  infinity with increasing $a$,
and the second one exists in a small region of negative $a$ with
an endpoint at ($a=-2-2b$, $T = 0$).  The slope of the critical
lines at $a=0$ is vertical (see inset of Fig. \ref{re-fig32}b).

As $b$ becomes positive, the reentrance on the positive side of
$a$ disappears (Fig. \ref{re-fig32}c), leaving only  phase III
(ferromagnetic).

\subsubsection{Model with one center} (Fig. \ref{re-fig29}d)

For this case, the quantities which satisfy Eq. (\ref{eq59}) are
given by
\begin{eqnarray}
X&= &\exp(4K_{2}) \nonumber \\
A&= &\exp(4K_{1})\cosh(4K_{3}) + \exp(-4K_{1})-2\cosh^{2}(2K_{3})\nonumber \\
B&= &\pm 2\{[\cosh(2K_{3}) +1]^{2} + [\exp(2K_{1})\cosh(2K_{3}) +
\exp(-2K_{1})]^{2}\}\nonumber \\
C&= &[\exp(2K_{1}) - \exp(-2K_{1})]^{2}\nonumber \\
A&= &\exp(4K_{1})\cosh(4K_{3}) + \exp(-4K_{1})+2\cosh^{2}(2K_{3})\nonumber \\
B&= &\pm 2\{[\cosh(2K_{3}) +1]^{2} - [\exp(2K_{1})\cosh(2K_{3}) -
\exp(-2K_{1})]^{2}\}\nonumber \\
C&= &[\exp(2K_{1}) + \exp(-2K_{1})]^{2}\label{eq62}
\end{eqnarray}

\begin{figure}[th]              
\centerline{\psfig{file=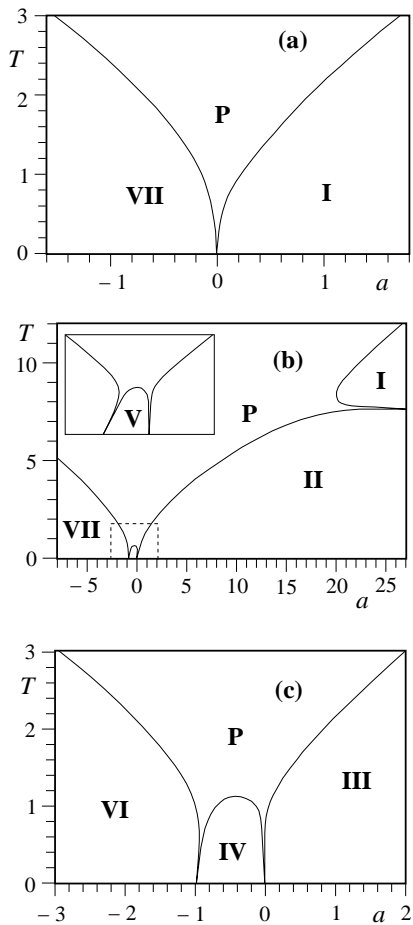,width=2.2in,angle=0}}
\vspace*{8pt} \caption{ One-center case: the same caption as that
of Fig. \ref{re-fig31} with (a) $b=-1.25$, (b) $b=-0.25$, (c)
$b=0.5$. \label{re-fig33}}
\end{figure}

The phase diagrams of this model shown in Fig. \ref{re-fig33} for
$b<-1$, $-1<b<0$ and $b>0$ are very similar to those of the
two-center model shown in Fig. \ref{re-fig32}. This is not
unexpected if one examines the ground state phase diagrams of the
two cases (Figs. 25b and 25c): their common point is the existence
of a partially disordered phase next to an ordered phase.  The
difference between the one- and two-center cases and the
three-center case shown above is that the latter has, in addition,
two boundaries, each of which separates two partially disordered
phases (see Fig. \ref{re-fig30}a). It is along these boundaries
that the two additional reentrances take place at finite $T$ in
the three-center case.

In conclusion of this subsection, we summarize that in  simple
models such as those shown in Fig. \ref{re-fig29}, we have found
two reentrant phases occurring on the temperature scale at a given
set of interaction parameters. A striking feature is the existence
of a reentrant phase between {\it two partially disordered phases}
which has not been found so far in any other model (we recall that
in other models, a reentrant phase is found between an ordered
phase and a partially disordered phase).

\subsection{Random-field aspects of the models}

Let us touch upon the random-field aspect of the model. A
connection between the Ising models presented above and the
random-field problem can be established. Consider for instance the
centered square lattice (Fig. \ref{re-fig17}). In the region of
antiferromagnetic ordering of sublattice 2 ($a=J_{2}/\mid
J_{1}\mid <-1$) , the spins on sublattice 1 (centered spins) are
free to flip. They act on their neighboring spins (sublattice 2)
as an annealed random field h. The probability distribution of
this random field at a site of sublattice 2 is given by
\begin{equation}
P(h)= \frac {1}{16}[6\delta (h)+4\delta (h+2J_{1})+ 4\delta
(h-2J_{1})+\delta (h+4J_{1})+ \delta (h-4J_{1})]\label{eq63}
\end{equation}
The random field at a number of spins is thus zero (diluted).
Moreover, this field distribution is somewhat correlated because
each spin of sublattice 1 acts on four spins of sublattice 2.
Since the spins on sublattice 1 are completely disordered at all
$T$, it is reasonable to consider this effective random-field
distribution as quenched. In addition to possible local annealed
effects, the phase transition at a finite $T$ of this model may be
a consequence of the above mentioned dilution and correlations of
the field distribution, because it is known that in 2D
random-field Ising model (without dilution) there is no such a
transition.\cite{Im}

The same argument is applied
to other
models studied above.

\section{Evidence of partial disorder and reentrance in other
frustrated systems}

The partial disorder\index{partial disorder} and the
reentrance\index{reentrance} which occur in exactly solved Ising
systems shown above are expected to occur also in models other
than the Ising one as well as in some three-dimensional systems.
Unfortunately, these systems cannot be exactly solved. One has to
use approximations or numerical simulations to study them. This
renders difficult the interpretation of the results. Nevertheless,
in the light of what has been found in exactly solved systems, we
can introduce the necessary ingredients into the model under study
if we expect the same phenomenon to occur.

As seen above, the most important ingredient for a partial
disorder and a reentrance to occur at low $T$ in the Ising model
is the existence of a number of free spins in the ground state.

In three dimensions, apart from a particular exactly solved
case\cite{Horiguchi} showing a reentrance,  a few Ising systems
such as the fully frustrated simple cubic
lattice,\cite{Blan,Diep85b} a stacked triangular Ising
antiferromagnet\cite{Blan85,Nagai} and a body-centered cubic (bcc)
crystal\cite{Aza89b} exhibit a partially disordered phase in the
ground state. We believe that reentrance should also exist in the
phase space of such systems though evidence is found numerically
only for the bcc case.\cite{Aza89b}

In two dimensions, a few non-Ising models show also evidence of a
reentrance. For the $q$-state Potts model,\index{Potts model}
evidence of a reentrance is found in a recent study of the
two-dimensional frustrated Villain lattice (the so-called piled-up
domino model)\index{Villain lattice} by a numerical transfer
matrix calculation\cite{foster1,foster2}. It is noted that the
reentrance occurs near the fully frustrated situation, i.e.
$\alpha_c=J_{AF}/J_F=-1$ (equal antiferromagnetic and
ferromagnetic bond strengths), for $q$ between $\simeq 1.0$ and
$\simeq 4$.  Note that there is no reentrance in the case $q=2$.
Below (above) this $q$ value, the reentrance occurs above (below)
the fully frustrated point $\alpha_c$ as shown in Figs. 30 and 31.
For $q$ larger than $\simeq 4$, the reentrance
disappears.\cite{foster2}

\begin{figure}[th]              
\centerline{\psfig{file=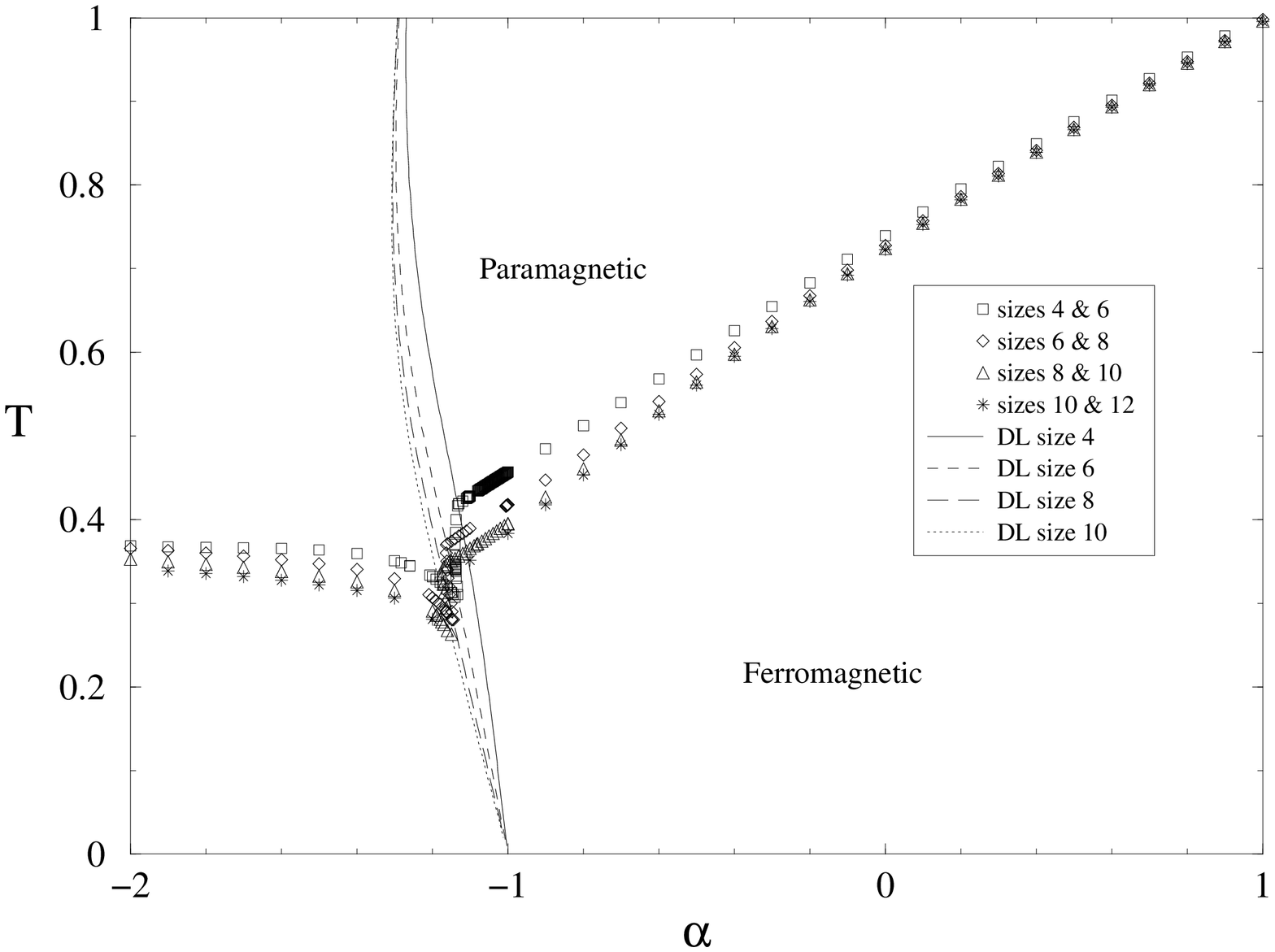,width=2.2in,angle=-90}}
\centerline{\psfig{file=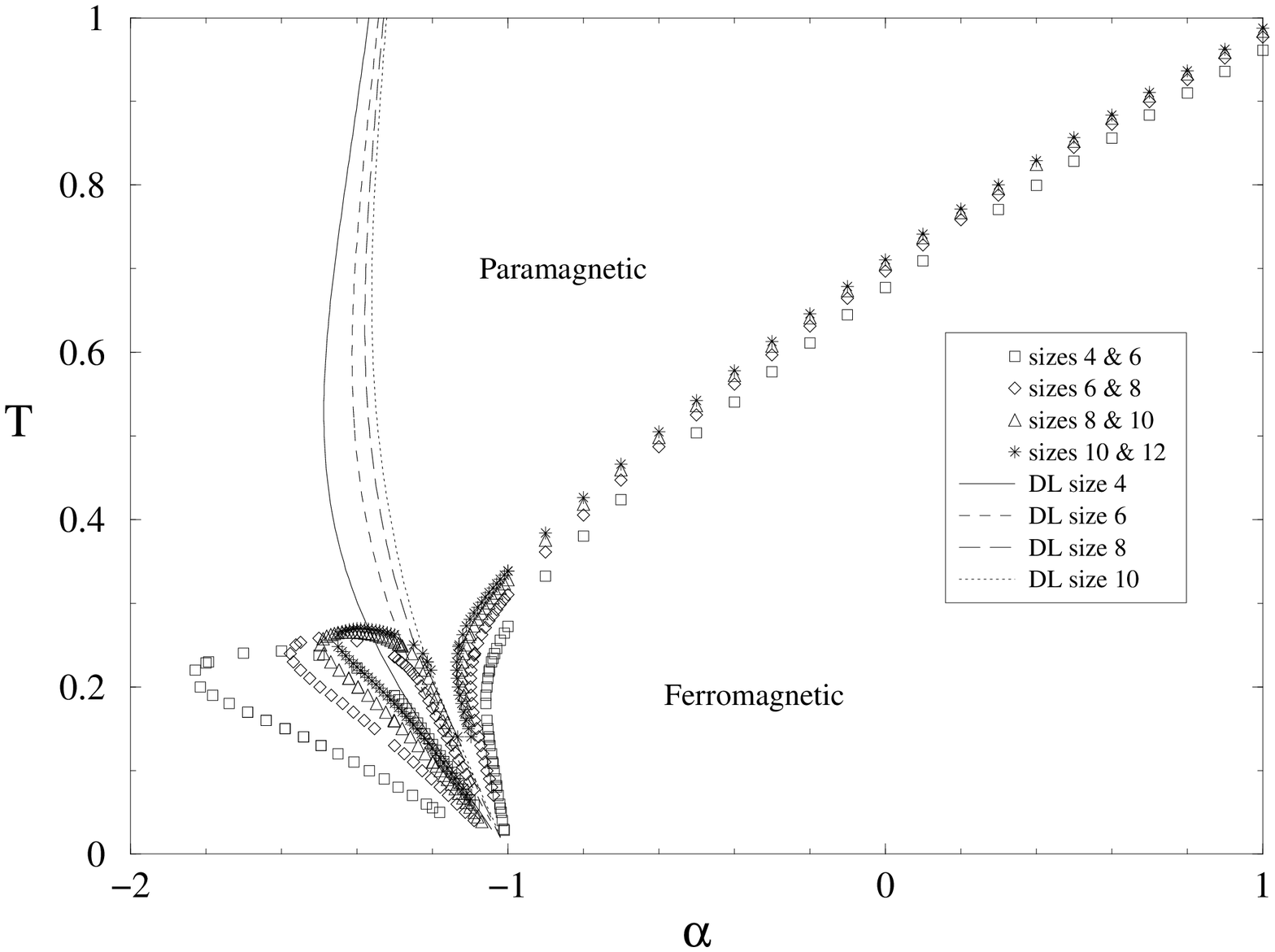,width=2.2in,angle=-90}}
\vspace*{8pt} \caption{Phase diagram for the Potts piled-up-domino
model with $q=3$:  periodic boundary conditions (top) and free
boundary conditions (bottom). The disorder lines\index{disorder
lines} are shown as lines, and the phase boundaries as symbols.
The numerical uncertainty is smaller than the size of the symbols
\cite{foster1}.}\label{re-fig34}
\end{figure}

\begin{figure}[th]
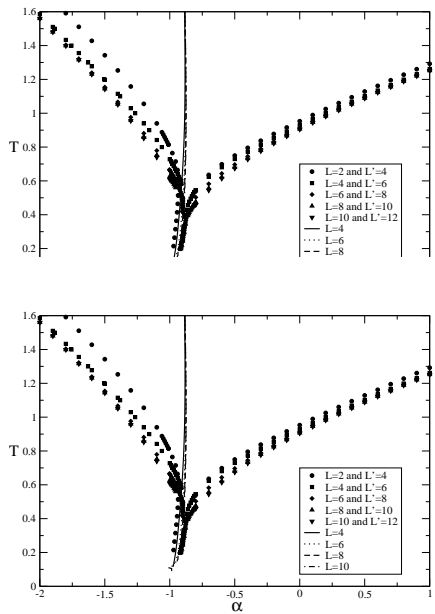
              
\centerline{\psfig{file=fig31a.eps,width=2.2in,angle=0}}
\centerline{\psfig{file=fig31a.eps,width=2.2in,angle=0}}
\vspace*{8pt} \caption{The phase diagram for $q=1.5$ found using
transfer matrices and the  phenomenological renormalization group
with periodic (top) and free boundary conditions (bottom).  The
points correspond to finite-size estimates for $T_c$, whilst the
lines correspond to the estimates for the disorder line,
($\alpha=J_2/J_1$) \cite{foster2}.}\label{re-fig35}
\end{figure}

A frustrated checkerboard lattice with XY spins shows also
evidence of a paramagnetic reentrance.\cite{Boubcheur}

\begin{figure}[th]              
\centerline{\psfig{file=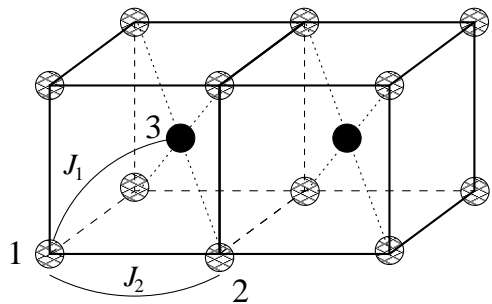,width=2.7in,angle=0}}
\vspace*{8pt} \caption{ bcc lattice. Spins are shown by gray and
black circles. Interaction between nn (spins numbered 1 and 3) is
denoted by $J_1$ and that between nnn (spins 1 and 2) by $J_2$.
Note that there is no interaction between black spins.
\label{re-fig36}}
\end{figure}

 In vector spin models such as the Heisenberg and XY models, the frustration is
shared by all bonds so that no free spins exist in the ground
state. However, one can argue that if there are several kinds of
local field in the ground state due to several kinds of
interaction, then there is a possibility that a subsystem with
weak local field is disordered at low $T$ while those of stronger
local field stay ordered up to higher temperatures.  This
conjecture has been verified in a number of recent works on
classical\cite{Santa1} and quantum spins\cite{Santa2,Quartu1}.
Consider for example Heisenberg spins $\bf{S}_i$ on a bcc lattice
with a unit cell shown in Fig. \ref{re-fig36}.\cite{Santa1}

\begin{figure}[th]              
\centerline{\psfig{file=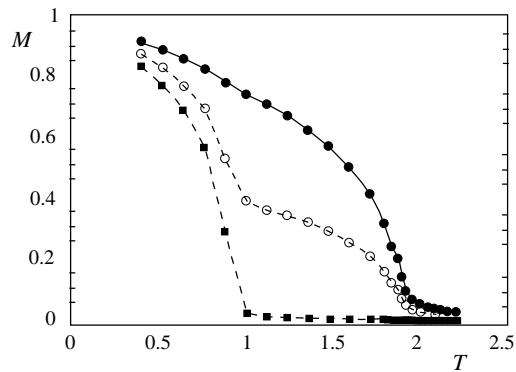,width=2.7in,angle=0}}
\vspace*{8pt} \caption{ Monte Carlo results for sublattice
magnetizations vs $T$ in the case $J_1=-1$, $J_2=-1.4$: black
squares and black circles are for sublattices 1 and 2,
respectively. Void circles indicate the total magnetization.
\label{re-fig37}}
\end{figure}

For convenience, let us call sublattice 1 the sublattice
containing the sites at the cube centers and sublattice 2 the
other sublattice. The Hamiltonian reads
\[\begin{array}{c}
 H=-\frac{1}{2}\sum_{<i,j>_1}{\bf S_{i}.S_{j}}-\frac{1}{2}
 \sum_{<i,j>_2}{\bf S_{i}.S_{j}}

\end{array}\]
where $\sum_{<i,j>_1}$ indicates the sum over the nn spin pairs
with exchange coupling $J_1$, while $\sum_{<i,j>_2}$ is limited to
the nnn spin pairs belonging to \mbox{sublattice 2} with exchange
coupling $J_2$.  It is easy to see that when $J_2$ is
antiferromagnetic the spin configuration is non collinear for
$J_2/|J_1|<-2/3$.  In the non collinear case, one can verify that
the local field acting a center spin (sublattice 1) is weaker in
magnitude than that acting on a corner spin (sublattice 2). The
partial disorder\index{partial disorder} is observed in Fig.
\ref{re-fig37}: the sublattice of black spins (sublattice 1) is
disordered at a low $T$.

\begin{figure}[th]              
\centerline{\psfig{file=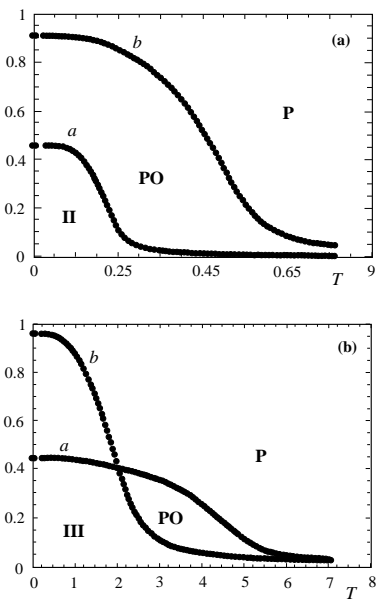,width=2.0in,angle=0}}
\vspace*{8pt} \caption{ Sublattice magnetizations vs $T$ in the
case $S_A=1/2$, $S_B=1$: (a) curve $a$ ($b$) is the
sublattice-1(2) magnetization in the case $J_{2A}/|J_1|=0.2$ and
$J_{2B}/|J_1|=0.9$ (b) curve $a$ ($b$) is the sublattice-1(2)
magnetization in the case $J_{2A}/|J_1|=2.2$ and
$J_{2B}/|J_1|=0.1$. P is the paramagnetic phase, PO the partial
order phase (only one sublattice is ordered), II and III are non
collinear spin configuration phases. See text for
comments.}\label{re-fig38}
\end{figure}

The same argument is applied for quantum spins.\cite{Quartu1}
Consider the bcc crystal as shown in Fig. \ref{re-fig36}, but the
sublattices are supposed now to have different spin magnitudes,
for example $S_A=1/2$ (sublattice 1) and $S_B=1$ (sublattice 2).
In addition, one can include nnn interactions in both sublattices,
namely $J_{2A}$ and $J_{2B}$. The Green function technique is then
applied for this quantum system.\cite{Quartu1} We show in Fig.
\ref{re-fig38} the partial disorder\index{partial disorder}
observed in two cases: sublattice 1 is disordered (Fig.
\ref{re-fig38}a) or sublattice 2 is disordered (Fig.
\ref{re-fig38}b). In each case, one can verify, using the
corresponding parameters, that in the ground state the spin of the
disordered sublattice has an energy lower than a spin in the other
sublattice.

\begin{figure}[th]              
\centerline{\psfig{file=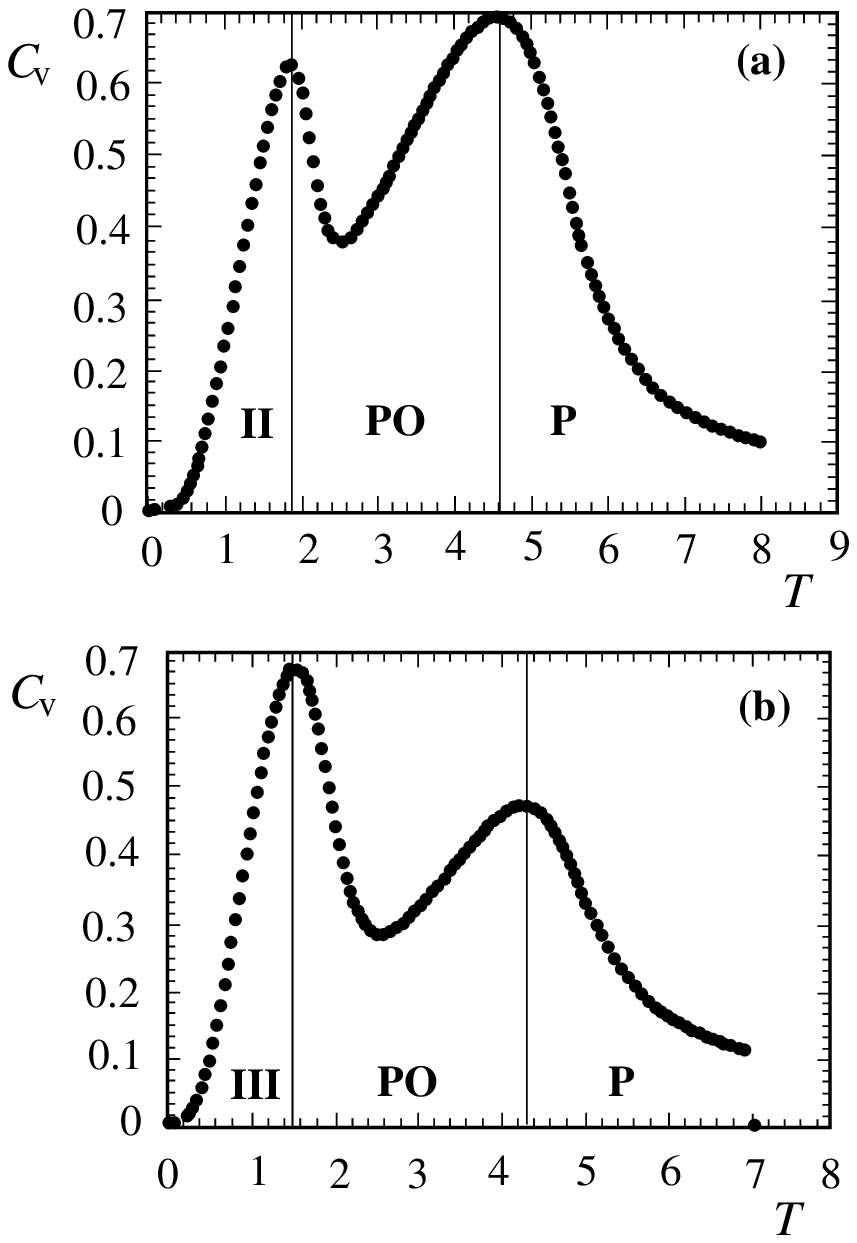,width=2.0in,angle=0}}
\vspace*{8pt} \caption{ Specific heat versus $T$ of the parameters
used in Fig. \ref{re-fig38}, $S_A=1/2$, $S_B=1$: (a)
$J_{2A}/|J_1|=0.2$ and $J_{2B}/|J_1|=0.9$ (b) $J_{2A}/|J_1|=2.2$
and $J_{2B}/|J_1|=0.1$. See the caption of Fig. \ref{re-fig38} for
the meaning of P, PO, II and III. See text for
comments.}\label{re-fig39}
\end{figure}

We show in Fig. \ref{re-fig39} the specific heat versus $T$ for
the parameters used in Fig. \ref{re-fig38}.  One observes the two
peaks corresponding to the two phase transitions associated with
the loss of sublattice magnetizations.

The necessary condition for the occurrence of a partial
disorder\index{partial disorder} at finite $T$ is thus the
existence of several kinds of site with different energies in the
ground state. This has been so far verified in a number of systems
as shown above.

\section{Conclusion}

In this chapter,  we have discussed some properties of
periodically frustrated Ising systems.
We have limited the discussion to exactly
solved models which possess at least a reentrant phase.  Other
Ising systems which involved approximations are discussed in
the chapter by Nagai et al (this book) and in the book by
Liebmann.\cite{Lieb}.

Let us emphasize that simple models having no bond disorder like
those presented in this chapter can possess complicated phase
diagrams due to the frustration generated by competing
interactions.  Many interesting physical phenomena such as
successive phase transitions, disorder lines, and reentrance are
found. In particular, a reentrant phase can occur in an infinite
region of parameters. For a given set of interaction parameters in
this region,  successive phase transitions take place on the
temperature scale, with one or two paramagnetic reentrant phases.

The relevance of disorder solutions\index{disorder solutions} for
the reentrance phenomena has also been pointed out. An interesting
finding is the occurrence of two disorder lines \index{disorder
lines} which divide the paramagnetic phase into regions of
different kinds of fluctuations (section 4.2). Therefore, care
should be taken in analyzing experimental data such as correlation
functions, susceptibility, etc. in the paramagnetic phase of
frustrated systems.

Although the reentrance is found in the models shown above by
exact calculations, there is no theoretical explanation why such a
phase can occur. In other words, what is the necessary and
sufficient condition for the occurrence of a reentrance?  We have
conjectured\cite{Aza87,Diep92} that the necessary condition for a
reentrance to take place is the existence of at least a partially
disordered phase next to an ordered phase or another partially
disordered phase in the ground state. The partial disorder is due
to the competition between different interactions.

The existence of a partial disorder\index{partial disorder} yields
the occurrence of a reentrance in most of known cases,\cite
{Vaks,Mo,Chi,Aza87,Diep91b,Diep92} except in some particular
regions of interaction parameters in the centered honeycomb
lattice (section 4.3.): the partial disorder alone is not
sufficient to make a reentrance as shown in Fig. \ref{re-fig27}a
and Fig. \ref{re-fig27}b, the finite zero-point entropy due to the
partial disorder of the ground state is the same for three cases
considered in Fig. \ref{re-fig26} , i.e. $S_{0}=\log (2)/3$ per
spin, but only one case yields a reentrance. Therefore, the
existence of a partial disorder is a necessary, but not
sufficient, condition for the occurrence of a reentrance.

The anisotropic character of the interactions can also favor the
occurrence of the reentrance. For example, the reentrant region is
enlarged by anisotropic interactions as in the centered square
lattice,\cite{Chi} and becomes infinite in the generalized
Kagom\'{e} model (section 4.2.). But again, this alone cannot
cause a reentrance as seen by comparing the anisotropic cases
shown in Fig. \ref{re-fig26}b and Fig. \ref{re-fig26}c: only in
the latter case a reentrance does  occur. The presence of a
reentrance may also require a coordination number at a disordered
site large enough to influence the neighboring ordered sites. When
it is too small such as in the case shown in Fig. \ref{re-fig26}b
(equal to two), it cannot induce a reentrance.  However, it may
have an upper limit to avoid the disorder contamination of the
whole system such as in the case shown in  Fig. \ref{re-fig27}a
where the coordination number is equal to six. So far, the 'right'
number is four in known reentrant systems shown above. Systematic
investigations of all possible ingredients are therefore desirable
to obtain a sufficient condition for the existence of a
reentrance.

Finally, let us emphasize that when a phase transition occurs
between states of different symmetries which have no special
group-subgroup relation, it is generally accepted that the
transition is of first order . However, the reentrance phenomenon
is a symmetry breaking alternative which allows one ordered phase
to change into another incompatible ordered phase by going through
an intermediate reentrant phase. A question which naturally arises
is under which circumstances does a system prefer an intermediate
reentrant phase to a first-order transition. In order to analyze
this aspect we have generalized the centered square lattice Ising
model into three dimensions.\cite{Aza89b} This is a special bcc
lattice. We have found that at low $T$ the reentrant region
observed in the centered square lattice shrinks into a first order
transition line which is  ended at a multicritical point from
which two second order lines emerge forming a narrow reentrant
region.\cite{Aza89b}

As a final remark, let us
mention that although the exactly solved systems
shown in this chapter are
models in statistical physics,
we believe that the results obtained in this work have qualitative
bearing on real frustrated magnetic systems.
In view of the
simplicity of these models,
we believe that the results found here will
have several applications in various areas of physics.

\section*{Note added for the third edition}
There is a number of papers dealing with exactly solved frustrated models published by J. Stre\u{c}ka and coworkers since 2006.  These models are essentially decorated Ising models in one or two dimensions.

In Ref. [53] ground-state and finite-temperature properties of the mixed spin-1/2 and spin-S Ising-Heisenberg diamond chains are examined within an exact analytical approach based on the generalized decoration-iteration map. A particular emphasis is laid on the investigation of the effect of geometric frustration, which is generated by the competition between Heisenberg- and Ising-type exchange interactions. It is found that an interplay between the geometric frustration and quantum effects gives rise to several quantum ground states with entangled spin states in addition to some semi-classically ordered ones. Among the most interesting results to emerge from our study one could mention a rigorous evidence for quantized plateaux in magnetization curves, an appearance of the round minimum in the thermal dependence of susceptibility times temperature data, double-peak zero-field specific heat curves, or an enhanced magneto-caloric effect when the frustration comes into play. The triple-peak specific heat curve is also detected when applying small external field to the system driven by the frustration into the disordered state.

In Ref. [54], the geometric frustration of the spin-1/2 Ising-Heisenberg model on the triangulated kagom\'e "triangles-in-triangles"lattice is investigated within the framework of an exact analytical method based on the generalized star-triangle mapping transformation. Ground-state and finite-temperature phase diagrams are obtained along with other exact results for the partition function, Helmholtz free energy, internal energy, entropy, and specific heat, by establishing a precise mapping relationship to the corresponding spin-1/2 Ising model on the kagom\'e lattice.  It  is  shown  that  the  residual  entropy  of  the  disordered  spin  liquid  phase  for  the  quantum  Ising-Heisenberg  model  is  significantly  lower  than  for  its  semiclassical  Ising  limit ($S_0/NTk_B= 0.2806$  and  0.4752, respectively), which implies that quantum fluctuations partially lift a macroscopic degeneracy of the ground-state manifold in the frustrated regime.

In Ref. [55], spin-1/2 Ising model with a spin-phonon coupling on decorated planar lattices partially amenable to lattice vibrations is examined using the decoration-iteration transformation and harmonic approximation. It is shown that the magneto-elastic coupling gives rise to an effective antiferromagnetic next-nearest-neighbor interaction, which competes with the nearest-neighbor interaction and is responsible for a frustration of decorating spins. A strong enough spin-phonon coupling consequently leads to an appearance of striking partially ordered and partially disordered phase, where a perfect antiferromagnetic alignment of nodal spins is accompanied with a complete disorder of decorating spins.

In the above works, the decorations are local couplings of an Ising spin to decorated Heisenberg spins or to phonons which can be summed up to renormalize the interactions between Ising spins. This leaves the systems solvable by star-triangle transformations, vertex models or other methods. Their results are interesting. We find again in these models striking features shown above in the present chapter for 2D solvable frustrated Ising models. In particular, partially disordered systems, multiple phase transitions and the reentrant phase due to the frustration are shown to exist in the phase diagrams.

For details, the reader is referred to Refs. [53,54,55].

%

\end{document}